\documentclass[twocolumn,showpacs,preprintnumbers,amsmath,amssymb]{revtex4-1}

\usepackage{graphicx,latexsym}
\usepackage{amsmath,amssymb,amsfonts}
\usepackage{bm}
\usepackage{braket}
\usepackage{mathrsfs}
\newcommand{\up}{\uparrow}
\newcommand{\down}{\downarrow}

\newcommand{\ep}{\varepsilon}
\def\lesssim{\ \raise.3ex\hbox{$<$}\kern-0.8em\lower.7ex\hbox{$\sim$}\ }
\def\gesim{\ \raise.3ex\hbox{$>$}\kern-0.8em\lower.7ex\hbox{$\sim$}\ }

\usepackage{ulem}
\usepackage{color}

\begin{document}
\title{Feasibility of FFLO superfluid Fermi atomic gas}
\author{Taira Kawamura\email{tairakawa@keio.jp}$^1$ and Yoji Ohashi$^1$}
\affiliation{Department of Physics, Keio University, 3-14-1 Hiyoshi, Kohoku-ku, Yokohama 223-8522, Japan}
\date{\today}
\begin{abstract}
We theoretically explore a promising route to achieve the Fulde-Ferrell-Larkin-Ovchinnikov (FFLO) state in a spin-imbalanced ultracold Fermi gas. In the current stage of cold atom physics, search for this exotic Fermi superfluid is facing two serious difficulties: One is the desperate destruction of the FFLO long-range order by FFLO pairing fluctuations, which precludes entering the phase through a second-order transition, even in three dimension. The other is the fierce competition with the phase separation into the BCS (Bardeen-Cooper-Schrieffer) state and the spin-polarized normal state. Including strong FFLO pairing fluctuations within the framework of the strong-coupling theory developed by Nozi\`eres and Schmitt-Rink, we show that the anisotropy of Fermi surface introduced by an optical lattice makes the FFLO state stable against the paring fluctuations. This stabilized FFLO state is also found to be able to overcome the competition with the phase separation under a certain condition. Since the realization of unconventional Fermi superfluids is one of the most exciting challenges in cold atom physics, our results would contribute to the further development of this field.
\end{abstract}
\maketitle
\par
\section{Introduction}
\par 
Recent successive experimental reports in condensed matter physics on the observation of the Fulde-Ferrell-Larkin-Ovchinnikov (FFLO) state \cite{Fulde1964,Larkin1964,Takada1969,Takada1970,Shimahara1994} in various materials, such as heavy fermion compounds CeCoIn$_5$ \cite{Bianchi2002,Bianchi2003,Kumagai2006} and CeCu$_2$Si$_2 $ \cite{Kitagawa2018}, organic conductor $\kappa$-(BEDT-TTF)$_2$Cu(NCS)$_2$ \cite{Singleton2000,Wright2011,Mayaffre2014}, as well as Fe-based superconductors KFe$_2$As$_2$ \cite{Cho2017} and FeSe \cite{Ok2020,Kasahara2020}, have stimulated the search for this exotic pairing state in cold Fermi gas physics \cite{Sheehy2007, Kinnunen2018, Hu2006, Parish2007, Chevy2010}. Since the FFLO state has also been discussed in superfluid liquid $^3$He under confinement \cite{Vorontsov2007,Aoyama2014, Wiman2016,Levitin2019,Shook2020}, chiral condensate in high-density QCD (quantum chromodynamics) system \cite{Alford2001,Casalbuoni2004,Anglani2014,Buballa2015}, nuclear matter (proton superconductors and neutron superfluids) \cite{Sedrakian2001,Isayev2002, Lee2018}, as well as non-equilibrium systems \cite{Doh2006,Zheng2015, Zheng2016,Huang2019,Kawamura2020,Kawamura2022}, once a FFLO superfluid Fermi atomic gas is realized, it is expected to be used as a quantum simulator for the study of these various systems, by using the high tunability of atomic gases. At present, although a consistent phase diagram with the presence of the FFLO state has been reported in a quasi-one-dimensional $^6$Li Fermi gas \cite{Liao2010}, there is no experimental evidence that the FFLO superfluid is really realized there. 
\par
The FFLO state is characterized by a spatially oscillating superfluid order parameter (which is symbolically written as $\Delta e^{i{\bm Q}\cdot{\bm r}}$), reflecting the non-zero center-of-mass momentum ${\bm Q}$ of the FFLO Cooper pair \cite{Fulde1964,Larkin1964,Takada1969,Takada1970,Shimahara1994}. Because of this unique pairing structure, the FFLO superfluid is weak against non-magnetic impurity scatterings, which is in contrast to the ordinary $s$-wave BCS (Bardeen-Cooper-Schrieffer) state (Anderson's theorem) \cite{Anderson}. Thus, as a difficulty of realizing the FFLO state, one needs to prepare a very clean sample. Indeed, the above-mentioned materials all satisfy this condition \cite{Bianchi2002,Bianchi2003,Kumagai2006,Kitagawa2018,Singleton2000,Wright2011,Mayaffre2014,Cho2017,Ok2020,Kasahara2020}. In addition, in metallic superconductivity, to stabilize the FFLO Cooper pairs with non-zero ${\bm Q}$, the misalignment of the Fermi surfaces between $\uparrow$-spin and $\downarrow$-spin electrons is tuned by the Zeeman effect under an external magnetic field above the Chandrasekhar-Clogston limit (where the ordinary uniform BCS state no longer exists) \cite{Clogston1,Clogston2}. However, such a strong magnetic field is known to also cause the orbital effect \cite{Shimahara2009}, which mixes the FFLO and the Abrikosov vortex states. This makes the realization of the {\it pure} FFLO state difficult, especially in three dimensions.
\par
Ultracold Fermi gases are free from the above-mentioned difficulties: (1) This system has no impurity. (2) The misalignment of two Fermi surfaces can be realized by simply imposing spin imbalance on the system, which is nothing to do with the unwanted orbital effect. Thus, at a glance, ultracold Fermi gases look very suitable for the FFLO state. 
\par
However, the search for the FFLO-superfluid Fermi gas in cold atom physics is also facing the following other serious problems: The first one is that the second-order phase transitions into both Fulde-Ferrell-type [$\Delta(\bm{r})=\Delta e^{i\bm{Q}\cdot \bm{r}}$] and Larkin-Ovchinikov-type [$\Delta(\bm{r})=\Delta \cos(\bm{Q}\cdot \bm{r})$] states are known to be completely suppressed by pairing fluctuations at non-zero temperatures \cite{Shimahara1998,Ohashi2002,Radzihovsky2009,Radzihovsky2011,Yin2014,Jakubczyk2017,Wang2018,Zdybel2021}. A similar instability phenomenon is also known in the ordinary BCS state in {\it one} and {\it two} dimensions (which is sometimes referred to as the Hohenberg-Mermin-Wagner theorem in the literature \cite{Hohenberg,Mermin}). However, in the FFLO case, it occurs even in {\it three} dimension \cite{note.crystal}. 
\par
This anomalous instability of the FFLO state originates from the spatial isotropy of an ultracold Fermi gas \cite{Shimahara1998, Ohashi2002, Radzihovsky2009, Radzihovsky2011, Yin2014, Jakubczyk2017, Wang2018, Zdybel2021}: When the system possesses such a continuous rotational symmetry in space, the FFLO state in this system is infinitely degenerate with respect to the direction of the FFLO ${\bm Q}$-vector. This degeneracy remarkably enhances FFLO pairing fluctuations, to completely destroy the FFLO long-range order \cite{note.q1D}. Roughly speaking, in the case of the ordinary BCS state at the superfluid phase transition temperature $T_{\rm c}$, effects $\Lambda_{\rm fluct}$ of low-energy pairing fluctuations on system properties are symbolically written as
\begin{equation}
\Lambda_{\rm fluct}\sim \int_0^{q_{\rm c}} q^{D-1}dq
{1 \over |{\bm q}|^2},
\label{eq.01}
\end{equation}
where $q_{\rm c}$ is a momentum cutoff. Equation (\ref{eq.01}) converges when the system dimension $D$ is larger than two, indicating the occurrence of the superfluid phase transition. In the FFLO state with ${\bm Q}\ne 0$, on the other hand, Eq. (\ref{eq.01}) is replaced by \cite{Ohashi2002}
\begin{equation}
\Lambda_{\rm fluct}\sim \int_0^{q_{\rm c}} q^{D-1}dq
{1 \over [|{\bm q}|-|{\bm Q}|]^2},
\label{eq.02}
\end{equation}
which always diverges irrespective of the system dimension $D$, leading to the vanishing superfluid phase transition. We briefly note that a similar instability is also discussed in inhomogeneous chiral condensations in high-density QCD \cite{Hidaka2015,Lee2015,Adhikari2017,Yoshiike2017,Pisarski2019,Ferrer2020}.
\par
The second problem is the phase separation \cite{Bedaque2003,Caldas2004,Cohen2005}. Indeed, instead of the desired FFLO state, phase separation into the BCS state and the spin-polarized normal state has so far only been observed in spin-imbalanced Fermi gases \cite{Shin2006,Zwierlein2006,Partridge2006,Shin2008,Mitra2016}.
\par
The purpose of this paper is to explore a promising route to reach the FFLO superfluid Fermi atomic gas, overcoming the above-mentioned two obstacles. To describe effects of strong FFLO pairing fluctuations, we extend the strong-coupling theory developed by Nozi\`eres and Schmitt-Rink (NSR) \cite{Nozieres1985} to the case with spin-imbalance. To remove the continuous rotational symmetry from a Fermi gas, we consider the case when the gas is loaded on a three-dimensional cubic optical lattice. We then examine how the FFLO state revives by the suppression of FFLO pairing fluctuations, when the continuous rotational symmetry of the system is replaced by the discrete four-fold one in the cubic optical lattice. Within the same strong-coupling scheme, we also examine whether or not the FFLO state that is stabilized by the optical lattice can also survive the competition with the phase separation. We briefly note that the importance of removing the continuous rotational symmetry from the system to stabilize the FFLO state has been pointed out \cite{Shimahara1998,Ohashi2002,Radzihovsky2009,Radzihovsky2011,Yin2014,Jakubczyk2017,Wang2018,Zdybel2021}; however, to our knowledge, the quantitative assessment of this idea, as well as the competition between the stabilized FFLO state and the phase separation, have not explicitly been examined yet.
\par
This paper is organized as follows. In Sec. II, we explain how to extend the NSR theory to a spin-imbalanced lattice Fermi gas. We show our results in Sec. III. Here, we first examine how the optical lattice stabilizes the FFLO state. We then study the competition between the stabilized FFLO state and the phase separation. Throughout this paper, we set $\hbar=k_{\rm B}=1$, and the system volume $V$ is taken to be unity, for simplicity.
\par
\section{Formulation}
\par
\subsection{Model Hamiltonian}
\par
We consider a two-component spin-imbalanced Fermi gas loaded on a three-dimensional cubic optical lattice. To model this system, we employ the attractive Hubbard Hamiltonian \cite{Tamaki2008, Kinnunen2018},
\begin{equation}
\hat{H} = -\sum_{i,j,\sigma} t_{i,j} \hat{c}^\dagger_{i,\sigma} \hat{c}_{j,\sigma} -U \sum_i \hat{n}_{i,\up} \hat{n}_{i,\down} 
-\sum_{i,\sigma} \mu_\sigma \hat{n}_{i,\sigma}.
\label{eq.H}
\end{equation}
Here, $\hat{c}_{i,\sigma}$ is the annihilation operator of a Fermi atom at the $i$-th lattice site, and $\hat{n}_{i,\sigma}=\hat{c}^\dagger_{i,\sigma} \hat{c}_{i,\sigma}$ is the number operator, where the pseudo-spin $\sigma=\up,\down$ describe two atomic hyperfine states. $-t_{i,j}$ is the hopping matrix element between the $i$-th and $j$-th sites. In this paper, the particle hopping is assumed to occur between the nearest-neighbor (NN) sites ($-t_{i,j}=-t$), as well as between the next nearest-neighbor (NNN) sites ($-t_{i,j}=-t'$). The on-site pairing interaction $-U (<0)$ is assumed to be tunable by adjusting the threshold energy of a Feshbach resonance \cite{Chin2010}. $\mu_\sigma$ is the Fermi chemical potential in the spin-$\sigma$ component. In this paper, we ignore effects of a harmonic trap, for simplicity \cite{note.trap,Meyrath2005,Es2010,Gaunt2013,Mukherjee2017}.
\par
For later convenience, we divide the model Hamiltonian in Eq. (\ref{eq.H}) into the sum $\hat{H} = \hat{H}_{\rm MF} +\hat{H}_{\rm FL}$ of the mean-field part $\hat{H}_{\rm MF}$ and the fluctuation part $\hat{H}_{\rm FL}$. The former has the form,
\begin{align}
\hat{H}_{\mathrm{MF}}
=
&-\sum_{i,j,\sigma} 
t_{i,j} \hat{c}_{i, \sigma}^{\dagger} \hat{c}_{j, \sigma}
-\sum_{i,\sigma} \tilde{\mu}_{\sigma} \hat{n}_{i, \sigma}
\notag\\
&
-\sum_i
\left[
\Delta_i \hat{c}^\dagger_{i, \up}\hat{c}^\dagger_{i, \down} 
+\Delta_i^* \hat{c}_{i,\down}\hat{c}_{i,\up} 
-{\Sigma_\up^{\rm H}\Sigma_\down^{\rm H} \over U}
-{\left|\Delta_i\right|^2 \over U}
\right].
\label{eq.Hmf}
\end{align}
Here, 
\begin{equation}
{\tilde \mu}_\sigma = \mu_\sigma -\Sigma_\sigma^{\rm H}
\label{eq.mu}
\end{equation}
is the effective Fermi chemical potential in the spin-$\sigma$ component, and 
\begin{align}
& \Delta_i = U \langle \hat{c}_{i,\down}\hat{c}_{i,\up}\rangle_{\rm MF},
\label{eq.OP.Delta}
\\[4pt]
& \Sigma_{\sigma}^{\rm H} = -U\langle \hat{n}_{i,-\sigma} \rangle_{\rm MF},
\label{eq.OP.rho}
\end{align}
are the superfluid order parameter and the Hartree energy, respectively. In Eqs. (\ref{eq.OP.Delta}) and (\ref{eq.OP.rho}), the average $\braket{\cdots}_{\rm MF}$ is taken for the mean-field Hamiltonian $\hat{H}_{\rm MF}$ in Eq.~\eqref{eq.Hmf}, and $-\sigma$ means the opposite spin component to $\sigma$. In this paper, we assume a uniform particle density, so that Eq. (\ref{eq.OP.rho}) has no site-dependence. For the superfluid order parameter $\Delta_i$ in Eq. (\ref{eq.OP.Delta}), we consider the the Fulde-Ferrell (FF) type \cite{Fulde1964,Takada1969,Takada1970,Shimahara1994},
\begin{equation}
\Delta_i=\Delta e^{i \bm{Q}\cdot \bm{R}_i},
\label{FF_order}
\end{equation}
where ${\bm R}_i$ is the spatial position of the $i$-th lattice site, and $\Delta$ is taken to be positive real, without loss of generality. 
\par
We note that Eq. (\ref{eq.Hmf}) has two kinds of mean-fields, that is, superfluid order parameter $\Delta_i$, and Hartree term $\Sigma_\sigma^{\rm H}$. (This type of theoretical framework is sometimes referred to the BCS-Stoner theory in the literature \cite{Dao2008,Kujawa2011,Cichy2014,Ptok2017}.) The reason for the appearance of the latter is that we are dealing with the single-band Hubbard model with a {\it finite band width}. In the case without optical lattice nor high-energy cutoff, the Hartree term is known to vanish \cite{Wyk2018, Ohashi2020},
\par
In momentum space, the mean-field Hamiltonian $\hat{H}_{\rm MF}$ in Eq. (\ref{eq.Hmf})  is written as, by using the FF superfluid order parameter in Eq. (\ref{FF_order}), 
\begin{align}
\hat{H}_{\mathrm{MF}}&=
\sum_{\bm{k}} \hat{\Psi}^\dagger_{\bm{k}}
\left[
\tilde{\xi}^{\rm s}_{\bm{k},\bm{Q}}\tau_3+
\tilde{\xi}^{\rm a}_{\bm{k},\bm{Q}}
-\Delta \tau_1
\right]
\hat{\Psi}_{\bm{k}} 
\notag\\
&\hspace{0.5cm}+
\sum_{\bm{k}}
\left[\tilde{\xi}_{-\bm{k}+\bm{Q}/2,\down} 
+{\Sigma_\up^{\rm H}\Sigma_\down^{\rm H} \over U}
+{\Delta^2 \over U} 
\right].
\label{eq.Hmf2}
\end{align}
Here, 
\begin{equation}
\hat{\Psi}_{\bm{k}}=
\begin{pmatrix}
\hat{c}_{\bm{k}+{\bm Q}/2,\uparrow} \\[4pt]
\hat{c}^\dagger_{-\bm{k}+{\bm Q}/2,\down} 
\end{pmatrix}
\end{equation}
is the two-component Nambu field \cite{Schrieffer}, and $\tau_i~(i=1,2,3)$ are  corresponding Pauli matrices acting on particle-hole space. In Eq. (\ref{eq.Hmf2}),
\begin{align}
\tilde{\xi}^{\rm s}_{\bm{k},\bm{Q}} &= 
\frac{1}{2}\big[\tilde{\xi}_{\bm{k}+\bm{Q}/2,\uparrow} +\tilde{\xi}_{-\bm{k}+\bm{Q}/2,\downarrow} \big],\\
\tilde{\xi}^{\rm a}_{\bm{k},\bm{Q}} &= 
\frac{1}{2}\big[\tilde{\xi}_{\bm{k}+\bm{Q}/2,\uparrow} -\tilde{\xi}_{-\bm{k}+\bm{Q}/2,\downarrow} \big],
\end{align}
where $\tilde{\xi}_{\bm{k},\sigma}=\ep_{\bm{k}} -\tilde{\mu}_\sigma$ is the kinetic energy of a Fermi atom, measured from the effective chemical potential $\tilde \mu_\sigma$ given in Eq. (\ref{eq.mu}). The single-particle energy $\ep_{\bm k}$ is given by
\begin{align}
\ep_{\bm{k}}= &-2t\sum_{\alpha=x,y,z} \cos(k_\alpha) 
-4t'\big[\cos(k_x)\cos(k_y)
\notag\\
&
+\cos(k_y)\cos(k_z)+\cos(k_x)\cos(k_z)\big].
\label{eq.band}
\end{align}
Here, the lattice constant is taken to be unity, for simplicity.
\par
The fluctuation part $\hat{H}_{\rm FL}$ of the Hamiltonian $\hat{H} = \hat{H}_{\rm MF} +\hat{H}_{\rm FL}$ has the form, in momentum space,
\begin{equation}
\hat{H}_{\rm FL} = -U \sum_{\bm{q}} \hat{\rho}_+(\bm{q}) \hat{\rho}_- (-\bm{q}).
\label{eq.HFL}
\end{equation}
Here, $\hat{\rho}_\pm(\bm{q}) = [\hat{\rho}_1(\bm{q}) \pm i\hat{\rho}_2(\bm{q})]/2$ are the generalized density operators \cite{Ohashi2003, Fukushima2007}, where
\begin{equation}
\hat{\rho}_{\alpha=1,2}(\bm{q}) =
\sum_{\bm{k}} \hat{\Psi}^\dagger_{\bm{k}+\bm{q}/2} \tau_\alpha  \hat{\Psi}_{\bm{k}-\bm{q}/2}
\end{equation}
physically describe amplitude ($\alpha=1$) and phase ($\alpha=2$) fluctuations of the superfluid order parameter around the mean-field value $\Delta$.
\par
\par
\subsection{Hartree-shifted Nozi\`eres-Schmitt-Rink (HNSR) theory}
\par
Following the standard NSR approach, we treat $\hat{H}_{\rm HF}$ in Eq. (\ref{eq.Hmf2}) as the non-perturbative Hamiltonian, to perturbatively include $\hat{H}_{\rm FL}$ in Eq. (\ref{eq.HFL}) within the Gaussian fluctuation level \cite{Ohashi2003,Fukushima2007,Melo1993,Randeria,Diener2008,He2015}. In the present case, one difference from the ordinary NSR case is that the non-perturbative part $\hat{H}_{\rm MF}$ already involves the Hartree corrections $\Sigma_\sigma^{\rm H}$. In this subsection, we explain how to construct this `Hartree-shifted' NSR (HNSR) theory for a spin-imbalance lattice Fermi gas.
\par
\subsubsection{Mean-field part}
\par
For the non-perturbative part $\hat{H}_{\rm MF}$, the corresponding $2\times 2$ matrix single-particle thermal Green's function has the form \cite{Schrieffer,Ohashi2003,Fukushima2007},
\begin{align}
{\bm G}({\bm k},i\omega_n)
&=
{1 
\over 
i\omega
-\tilde{\xi}^{\rm a}_{\bm{k},\bm{Q}}
-\tilde{\xi}^{\rm s}_{\bm{k},\bm{Q}}\tau_3
+\Delta \tau_1
}
\notag\\[4pt]
&=
{
[i\omega-\tilde{\xi}^{\rm a}_{\bm{k},\bm{Q}}]
+\tilde{\xi}^{\rm s}_{\bm{k},\bm{Q}}\tau_3
-\Delta \tau_1
\over
[i\omega-\tilde{\xi}^{\rm a}_{\bm{k},\bm{Q}}]^2-{\tilde E}_{{\bm k},{\bm Q}}
},
\label{eq.Green}
\end{align}
where $E_{{\bm k},{\bm Q}}=\sqrt{({\tilde \xi}_{{\bm k},{\bm Q}}^{\rm s})^2+\Delta^2}$, and $\omega_n$ is the fermion Matsubara frequency. Using Eq. (\ref{eq.Green}), we can evaluate the magnitude $\Delta$ of the FF superfluid order parameter in Eq. (\ref{FF_order}), as well as the the Hartree energies $\Sigma_\sigma^{\rm H}$ in Eq. (\ref{eq.OP.rho}), as, respectively,
\begin{align}
\Delta
&=
UT\sum_{{\bm k},\omega_n}G_{12}({\bm k},i\omega_n)
\nonumber
\\
&=
U\sum_{\bm{k}} \frac{\Delta}{2E_{\bm{k},\bm{Q}}}
\left[
1 -f(E^+_{\bm{k},\bm{Q}}) -f(E^-_{\bm{k},\bm{Q}})
\right],
\label{eq.gap}
\\[4pt]
\Sigma_\up^{\rm H}
&=
UT\sum_{{\bm k},\omega_n}G_{22}({\bm k},i\omega_n)e^{-i\omega_n\delta}
\nonumber
\\
&=
\scalebox{0.9}{$\displaystyle
-{U \over 2}
\sum_{\bm{k}}
\left[
\left[1+
{
{\tilde \xi}_{{\bm k},{\bm Q}}^{\rm s} 
\over 
E_{{\bm k},{\bm Q}}
}
\right] 
f(E^-_{\bm{k},\bm{Q}})
+
\left[1-
{
{\tilde \xi}_{{\bm k},{\bm Q}}^{\rm s} 
\over 
E_{{\bm k},{\bm Q}}
}
\right]
f(-E^+_{\bm{k},\bm{Q}}) 
\right] $}
\nonumber
\\
&\equiv 
-U\sum_{\bm k}n_{-{\bm k}+{\bm Q}/2,\down}^{\rm MF},
\label{eq.MF.density.up}
\\[4pt]
\Sigma_\down^{\rm H}
&=
-UT\sum_{{\bm k},\omega_n}G_{11}({\bm k},i\omega_n)e^{i\omega_n\delta}
\nonumber
\\
&=
\scalebox{0.9}{$\displaystyle
-{U \over 2}
\sum_{\bm{k}}
\left[
\left[1+
{
{\tilde \xi}_{{\bm k},{\bm Q}}^{\rm s} 
\over 
E_{{\bm k},{\bm Q}}
}
\right] 
f(E^+_{\bm{k},\bm{Q}})
+
\left[1-
{
{\tilde \xi}_{{\bm k},{\bm Q}}^{\rm s} 
\over 
E_{{\bm k},{\bm Q}}
}
\right]
f(-E^-_{\bm{k},\bm{Q}}) 
\right] $}
\nonumber
\\
&\equiv 
-U\sum_{\bm k}n_{{\bm k}+{\bm Q}/2,\up}^{\rm MF}.
\label{eq.MF.density.down}
\end{align}
Here, $f(\pm E_{{\bm k},{\bm Q}}^\pm)$ is the Fermi distribution function, and $\delta$ an infinitesimally small positive number. $E^\pm_{\bm{k},\bm{Q}}=E_{\bm{k},\bm{Q}} \pm \tilde{\xi}^{\rm a}_{\bm{k},\bm{Q}}$ describes Bogoliubov single-particle excitations in the FF state \cite{Fulde1964,Larkin1964,Takada1969,Takada1970,Shimahara1994}. $n^{\rm MF}_{{\bm k}+{\bm Q}/2,\up}=\langle c^\dagger_{{\bm k}+{\bm Q}/2,\up}c_{{\bm k}+{\bm Q}/2,\up}\rangle_{\rm MF}$ and $n^{\rm MF}_{-{\bm k}+{\bm Q}/2,\down}=\langle c^\dagger_{-{\bm k}+{\bm Q}/2,\down}c_{-{\bm k}+{\bm Q}/2,\down}\rangle_{\rm MF}$ are the mean-field momentum distributions of Fermi atoms in the spin-$\up$ and the spin-$\down$ components, respectively.
\par
The ${\bm Q}$-vector in Eq. (\ref{FF_order}), which physically means the center-of-mass momentum of a FF Cooper pair, is determined from the stationary condition, 
\begin{equation}
{\partial\Omega_{\rm MF} \over \partial{\bm Q}}=0,
\label{eq.Q}
\end{equation}
where
\begin{align}
\Omega_{\rm MF}
&=
-T\ln\big[{\rm Tr}[e^{-\beta \hat{H}_{\rm MF}}]\big]
\nonumber
\\
=&
-\sum_{\bm{k}}\bigg[ \tilde{\xi}_{-\bm{k}+\bm{Q}/2,\down} -E^-_{\bm{k},\bm{Q}} -T\Big[ \ln\big(1+e^{-\beta E^+_{\bm{k},\bm{Q}}}\big) 
\notag\\
&
+\ln\big(1+e^{-\beta E^-_{\bm{k},\bm{Q}}}\big) \Big] 
+{\Sigma_\up^{\rm H}\Sigma_\down^{\rm H} \over U}
+\frac{\Delta^2}{U}\bigg]
\label{eq.Omega.MF}
\end{align}
is the mean-field thermodynamic potential. Equation (\ref{eq.Q}) gives the vanishing total current condition \cite{Takada1969},
\begin{equation}
\scalebox{0.95}{$\displaystyle
\bm{J}=\sum_{\bm k}
\left[
\Big[{\bm k}+{{\bm Q} \over 2}\Big]n^{\rm MF}_{{\bm k}+{\bm Q}/2,\up}
+
\Big[-{\bm k}+{{\bm Q} \over 2}\Big]n^{\rm MF}_{-{\bm k}+{\bm Q}/2,\down}
\right]
=0$},
\label{eq.Bloch_A}
\end{equation}
which is consistent with the so-called Bloch's theorem, stating the vanishing spontaneous total current in any thermodynamically stable state \cite{Bohm1949,Ohashi1996}. 
\par
We briefly note that the gap equation (\ref{eq.gap}), as well as the Hartree corrections in Eqs. (\ref{eq.MF.density.up}) and (\ref{eq.MF.density.down}), are also obtained from the stationary conditions, respectively,
\begin{align}
&
{\partial\Omega_{\rm MF} \over \partial\Delta}=0,
\label{stationaryA}
,\\[4pt]
&
{\partial\Omega_{\rm MF} \over \partial\Sigma_\sigma^{\rm H}}=0.
\label{stationaryB}
\end{align}
\par
\begin{figure}[t]
\centering
\includegraphics[width=8.3cm]{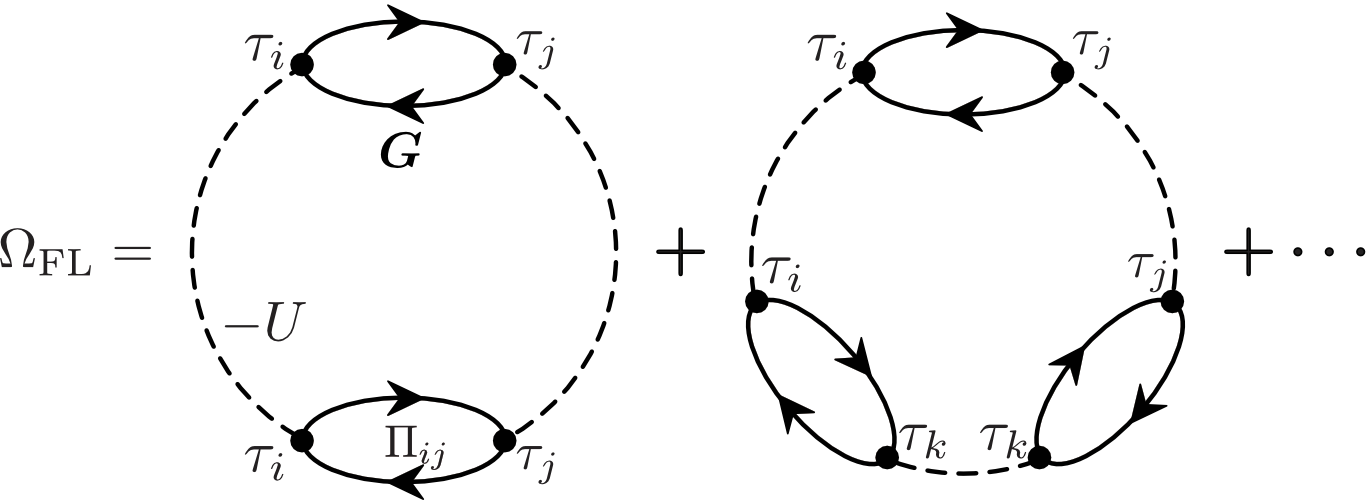}
\caption{Fluctuation correction $\Omega_{\rm FL}$ to the thermodynamic potential $\Omega$ in the HNSR theory. The solid line and the dashed line are the $2\times 2$ matrix mean-field single-particle thermal Green's function in Eq. (\ref{eq.Green}), and the pairing interaction $-U$, respectively. $\Pi_{ij}$ is the pair-correlation function in Eq. (\ref{eq.Pi_ij}). The solid circle denotes a Pauli matrix $\tau_i$ in the Nambu representation.
}
\label{fig1} 
\end{figure}
\par
\subsubsection{Fluctuation corrections to thermodynamic potential $\Omega$}
\par
We treat the fluctuation term $\hat{H}_{\rm FL}$ in Eq. (\ref{eq.HFL}) within the NSR diagrams shown in Fig. \ref{fig1} \cite{Nozieres1985,Ohashi2003,Fukushima2007,Melo1993,Randeria,Diener2008,He2015}, to evaluate fluctuation correction $\Omega_{\rm FL}$ to the thermodynamic potential $\Omega=\Omega_{\rm MF}+\Omega_{\rm FL}$. In Fig. \ref{fig1}, since the present single-particle thermal Green's function ${\bm G}$ in Eq. (\ref{eq.Green}) already involves the Hartree self-energy $\Sigma_\sigma^{\rm H}$ given in Eqs. (\ref{eq.MF.density.up}) and (\ref{eq.MF.density.down}), the diagrammatic series starts from the second order term in terms of $\hat{H}_{\rm FL}$ in order to avoid double counting \cite{Tajima2019}. The summation of these diagrams gives
\begin{equation}
\Omega_{\rm FL} = \frac{T}{2} \sum_{\bm{q},i\nu_m} {\rm Tr}
\left[
\ln 
\bigl[
1 +U{\bm{\Pi}}(\bm{q},i\nu_m)
\bigr] -U{\bm{\Pi}}(\bm{q},i\nu_m)
\right].
\label{eq.Omega.FL}
\end{equation}
Here, $\nu_m$ is the boson Matsubara frequency, and
\begin{equation}
{\bm{\Pi}}(\bm{q},i\nu_m)=
\begin{pmatrix}
\Pi_{-+}(\bm{q},i\nu_m) &\Pi_{--}(\bm{q},i\nu_m)  \\[4pt]
\Pi_{++}(\bm{q},i\nu_m) &\Pi_{+-}(\bm{q},i\nu_m) 
\end{pmatrix}
\end{equation}
is the 2$\times$2 matrix pair correlation function, where
\begin{align}
&\Pi_{\pm\pm}(\bm{q},i\nu_m)
\notag\\
&\hspace{0.5cm}=
T\sum_{\bm{k},i\omega_n}{\rm Tr}\big[\tau_\pm \bm{G}(\bm{k}+\bm{q},i\omega_n+i\nu_m) \tau_\pm \bm{G}(\bm{k},i\omega_n)\big],
\label{eq.Pi_ij}
\end{align}
with $\tau_{\pm}=[\tau_1 \pm i \tau_2]/2$. In Eq. (\ref{eq.Omega.FL}), the last term in $[\cdot\cdot\cdot]$ removes the first-order contribution of $\hat{H}_{\rm FL}$ from $\Omega_{\rm FL}$.
\par
The HNSR equation for the number $N_\sigma$ of Fermi atoms in the spin-$\sigma$ component is derived from the thermodynamic identity,
\begin{equation}
N_\sigma
=-{\partial \Omega \over \partial \mu_\sigma}
=-{\partial \Omega_{\rm MF} \over \partial \mu_\sigma}
-{\partial \Omega_{\rm FL} \over \partial \mu_\sigma}.
\label{eq.num}
\end{equation}
The set of Eqs. (\ref{eq.gap})-(\ref{eq.MF.density.down}), (\ref{eq.Bloch_A}) and (\ref{eq.num}), is the basis of the HNSR theory. 
\par
In considering a spin-imbalanced lattice Fermi gas, it is convenient to introduce the polarization,
\begin{equation}
P={N_\uparrow-N_\downarrow \over N_\uparrow+N_\downarrow},
\label{eq.polarization}
\end{equation}
and the filling fraction,
\begin{equation}
n={N \over M},
\label{eq.filling}
\end{equation}
where $N=N_\up+N_\down$ and $M$ are the total number of Fermi atoms and the number of lattice sites, respectively. Since we assume uniform particle density in this paper, the filling fraction $n$ equals the number of occupied Fermi atoms at each lattice site.
\par
In the superconducting case, on the other hand, the population imbalance is tuned by adjusting an external magnetic field. To describe this situation, useful parameters are the effective `magnetic field' $h=[\mu_\up-\mu_\down]/2$, and the averaged Fermi chemical potential $\mu=[\mu_\up+\mu_\down]/2$.
\par
As in the NSR theory \cite{Ohashi2003,Diener2008,He2015}, the present HNSR theory also satisfies the required Goldstone theorem, stating the existence of the gapless collective phase oscillation in the superfluid phase. We briefly note that a similar extension of the NSR theory to a spin-imbalance Fermi gas has been discussed in Ref. \cite{Tempere2008}, where the phase separation in a trapped spin-imbalanced Fermi gas is successfully described by the combined NSR theory with the local density approximation.
\par
\begin{figure*}[t]
\centering
\includegraphics[width=13.5cm]{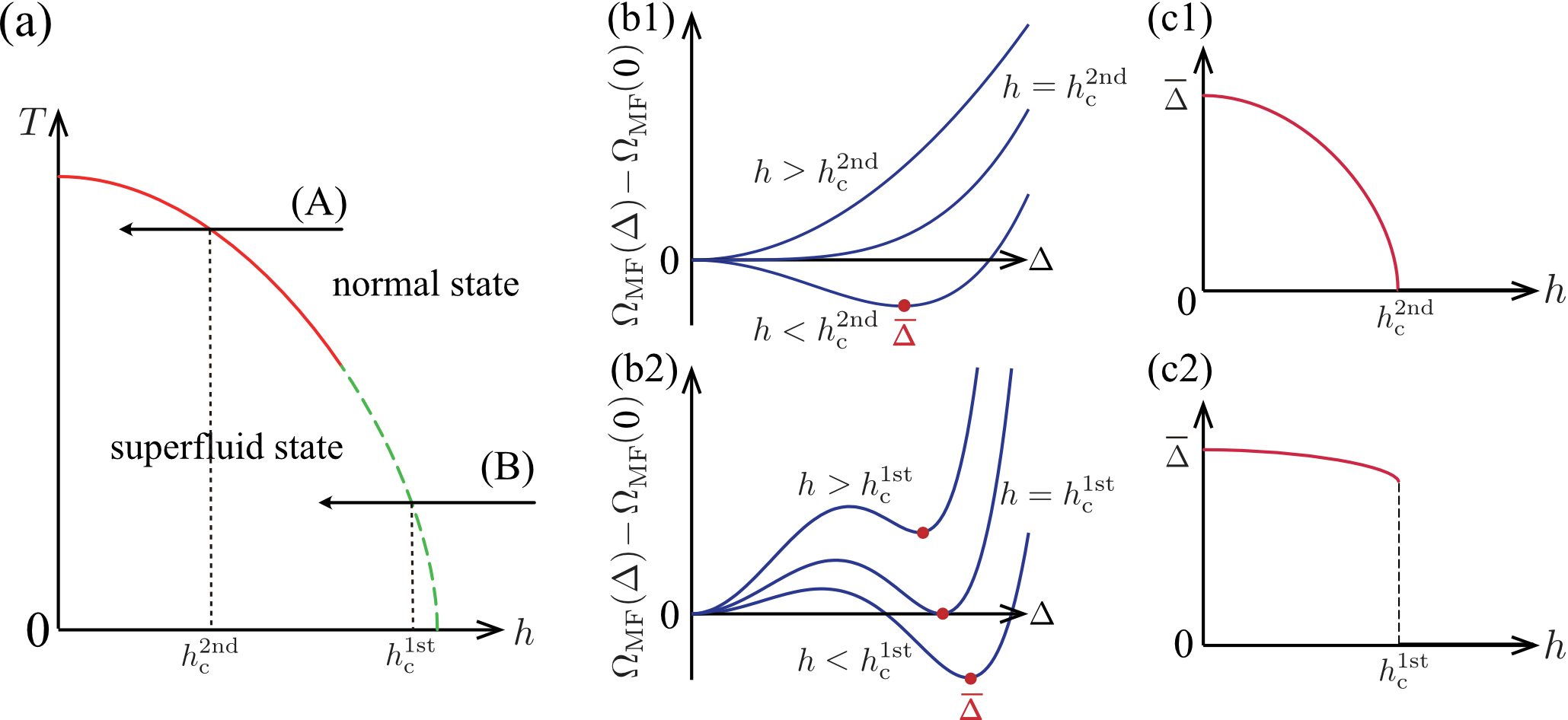}
\caption{(a) Schematic superfluid phase diagram to explain how to determine phase boundaries. The solid (dashed) line represents an assumed second-order (first-order) superfluid phase transition temperature. When we approach the superfluid phase (BCS or FFLO state) from the normal state by decreasing the effective magnetic field $h=[\mu_\up-\mu_\down]/2$ along the path-(A) (path-(B)), and passes through the second-order (first-order) phase transition line under the condition that the averaged chemical potential $\mu=[\mu_\up+\mu_\down]/2$ is fixed, the thermodynamic potential $\Omega_{\rm MF}(\Delta)$ as a function of $\Delta$ varies as shown in panel (b1) (panel (b2)). In this case, as shown in panel (c1) (panel (c2)), the superfluid order parameter ${\bar \Delta}$ continuously (discontinuously) grows from zero in the superfluid phase below $h=h_{\rm c}^{\rm 2nd}$ ($h=h_{\rm c}^{\rm 1st}$). In panels (b1) and (b2), the solid circle represents the value of $\Delta~(>0)$ at which $\Omega_{\rm MF}(\Delta)$ takes a local minimum. In the normal state, $\Omega_{\rm MF}(\Delta)$ always the smallest at $\Delta=0$.}
\label{fig2} 
\end{figure*}
\par
\subsection{Determination of phase boundaries}
\par
\subsubsection{First- and second-order phase transition in $T$-$h$ phase diagram}
\par
We explain how to determine the boundary between the superfluid (BCS or FFLO) state and the normal state, by using the schematic $T$-$h$ phase diagram in Fig. \ref{fig2}(a). When one passes through the second-order superfluid phase transition along the path-(A), the HNSR theory (as well as the NSR theory) determines the superfluid order parameter $\Delta$ so as to minimize $\Omega_{\rm MF}(\Delta)$ in Eq. (\ref{eq.Omega.MF}). Around the critical magnetic field $h=h_{\rm c}^{\rm 2nd}$, $\Omega_{\rm MF}(\Delta)$ behaves like Fig. \ref{fig2}(b1), so that the superfluid order parameter ${\bar \Delta}$ at the minimum of $\Omega_{\rm MF}$ continuously grows from zero below $h_{\rm c}^{\rm 2nd}$. According to the Ginzburg-Landau theory, when one expands $\Omega_{\rm MF}$ with respect to $\Delta$, the coefficient of the second-order term changes its sign at the second-order phase transition. Thus, $h_{\rm c}^{\rm 2nd}$ is determined so as to satisfy
\begin{equation}
0=\left.
{\partial^2 \Omega_{\rm MF}(\Delta}) \over \partial\Delta^2
\right|_{\Delta\to 0}
={1 \over U}-\Pi({\bm q}={\bm Q},i\nu_m=0,\mu,h_{\rm c}^{\rm 2nd}),
\label{eq.2nd_gap}
\end{equation}
where
\begin{equation}
\Pi(\bm{q},i\nu_m,\mu,h)
=
\sum_{\bm{k}} 
{
1
-f(\tilde{\xi}_{\bm{k}+\bm{q}/2,\uparrow})
-f(\tilde{\xi}_{-\bm{k}+\bm{q}/2,\downarrow})
\over
\tilde{\xi}_{\bm{k}+\bm{q}/2,\uparrow} 
+\tilde{\xi}_{-\bm{k}+\bm{q}/2,\downarrow}-i\nu_m
}
\label{eq.Pi.N}
\end{equation}
is the pair-correlation function in the normal state. The Hartree term involved in ${\tilde \xi_{{\bm p},\sigma}}$ is reduced to, at $h_{\rm c}^{\rm 2nd}$,
\begin{equation}
\Sigma_\sigma^{\rm H}=-U\sum_{\bm k}f({\tilde \xi}_{{\bm k},-\sigma}).
\label{Hatree_N}
\end{equation}
Equation (\ref{eq.2nd_gap}) is essentially the same as the gap equation (\ref{eq.gap}) in the limit $\Delta\to+0$. When ${\bm Q}=0$, Eq. (\ref{eq.2nd_gap}) is reduced to the ordinary BCS gap equation at $T_{\rm c}$ under an external magnetic field. We briefly note that, in the NSR theory, this $T_{\rm c}$-equation is usually derived from the Thouless criterion, stating that the superfluid phase transition occurs when the particle-particle scattering matrix,
\begin{equation}
\Gamma({\bm q},i\nu_m)\equiv -{U \over 1-U\Pi({\bm q},i\nu_n)},
\label{eq.gamma}
\end{equation}
has a pole at ${\bm q}=\nu_m=0$ \cite{Thouless1960,Liu2006,Wardh2018,Durel2020}.
\par
In the FF case (${\bm Q}\ne 0$), the vanishing current condition in Eq. (\ref{eq.Bloch_A}) is automatically satisfied at the second-order superfluid phase transition, regardless of the value of ${\bm Q}$:
\begin{align}
\bm{J}(\Delta=0)=
&
\sum_{\bm k}
\bigg[
\Big[{\bm k}+{{\bm Q} \over 2}\Big]f(\xi_{{\bm k}+{\bm Q}/2,\up})
\notag\\
&
+
\Big[-{\bm k}+{{\bm Q} \over 2}\Big]f(\xi_{-{\bm k}+{\bm Q}/2,\down})
\bigg]
\nonumber
\\
=&
\sum_{{\bm k},\sigma}{\bm k}f(\xi_{{\bm k},\sigma})
=0.
\label{eq.Bloch}
\end{align}
Thus, the FF ${\bm Q}$-vector at the phase boundary is determined from Eq. (\ref{eq.2nd_gap}) so that the largest $h_{\rm c}^{\rm 2nd}$ can be obtained.
\par
We numerically solve Eq. (\ref{eq.2nd_gap}), together with the HNSR number equation (\ref{eq.num}) with $\Delta=0$, to determine $({\bm Q},h_{\rm c}^{\rm 2nd},\mu)$ (or equivalently $({\bm Q},\mu_\up,\mu_\down)$) at the second-order superfluid phase transition, for a given parameter set $(T,U,N_\up,N_\down)$. The latter equation is obtained from Eqs. (\ref{eq.Omega.MF}), (\ref{eq.Omega.FL}), and Eq. (\ref{eq.num}), as, by setting $\Delta=0$,
\begin{align}
N_\sigma=&
\sum_{\bm k}f({\tilde \xi}_{{\bm k},\sigma})-
T\sum_{\sigma'=\up,\down} \sum_{{\bm q}, i\nu_m}
e^{i\nu_m\delta}
\left[\Gamma({\bm q},i\nu_m)+U \right]
\notag\\
&\times
\left(
{\partial \Pi({\bm q},i\nu_m) \over \partial {\tilde \mu}_{\sigma'}}
\right)_{T,{\tilde \mu}_{-\sigma'}}
\left(
{\partial {\tilde \mu}_{\sigma'} \over \partial \mu_\sigma}
\right)_{T,{\tilde \mu}_{-\sigma'}},
\label{eq.num_normal}
\end{align}
where $\Pi({\bm q},i\nu_m)$ is given in Eq. (\ref{eq.Pi.N}), and 
\begin{eqnarray}
\left(
\begin{array}{cc}
{\partial {\tilde \mu}_\up \over \partial\mu_\up} &
{\partial {\tilde \mu}_\down \over \partial\mu_\up} \\[4pt]
{\partial {\tilde \mu}_\up \over \partial\mu_\down} &
{\partial {\tilde \mu}_\down \over \partial\mu_\down} 
\end{array}
\right)
=
{1 \over 1-U^2\kappa_\up\kappa_\down}
\left(
\begin{array}{cc}
1 & U\kappa_\up \\[4pt]
U\kappa_\down & 1
\end{array}
\right).
\label{eq.num_FL}
\end{eqnarray}
In Eq. (\ref{eq.num_FL}), 
\begin{equation}
\kappa_{\sigma}=
\sum_{\bm k}
{
\partial f({\tilde \xi}_{{\bm k},\sigma}) \over \partial {\tilde \mu}_\sigma
}=
{1 \over 4T}\sum_{\bm k}{\rm sech}^2
\left({{\tilde \xi}_{{\bm k},\sigma} \over 2T}\right)
\end{equation}
is the isothermal compressibility in the mean-field approximation, which comes from the Hartree term involved in ${\tilde \mu}$ in Eq. (\ref{eq.mu}).
\par
We note that the coupled equations (\ref{eq.2nd_gap}) with (\ref{eq.num_normal}) may give {\it multiple} $\bm{Q}_l$-vectors $(l=1,2,\cdot\cdot\cdot)$ having the same magnitude $|{\bm Q}_1|=|{\bm Q}_2|=\cdot\cdot\cdot$, reflecting the discrete four-fold rotational symmetry of the cubic optical lattice. Thus, although we have assumed the `single-${\bm Q}$' FF state in Eq. (\ref{FF_order}), a more complicated FFLO-type superfluid order parameter, for example,
\begin{equation}
\Delta_i= \Delta \sum_l e^{i\bm{Q}_l\cdot \bm{R}_i},
\label{eq.OP.FFLO}
\end{equation}
is also possible \cite{note2,Shimahara2}. However, at the second-order superfluid transition ($\Delta=0$), any multiple-${\bm Q}$ FFLO state has the same critical magnetic field $h_{\rm c}^{\rm 2nd}$ as that in the single-${\bm Q}$ case. This means that the coupled equations (\ref{eq.2nd_gap}) with (\ref{eq.num_normal}) can cover all non-uniform superfluid states with the superfluid order parameter in Eq \eqref{eq.OP.FFLO}. Keeping this in mind, we simply call the superfluid state with ${\bm Q}\ne 0$ the FFLO state in what follows, as far as the second-order phase transition of this state is considered.
\par
We next consider the first-order critical magnetic field $h_{\rm c}^{\rm 1st}$, which we meet when moving along the path-(B) in Fig. \ref{fig2}(a) under the condition that that averaged chemical potential $\mu$ is fixed. Near the first-order phase transition, $\Omega_{\rm MF}(\Delta)$ exhibits a double-minimum structure \cite{Gubbels2013}, as schematically shown in Fig. \ref{fig2}(b2). In the HNSR scheme, $h_{\rm c}^{\rm 1st}$ is determined from the condition,
\begin{equation}
\Omega_{\rm MF}(\Delta=0,\mu,h_{\rm c}^{\rm 1st},T)=
\Omega_{\rm MF}({\bar \Delta}>0,\mu,h_{\rm c}^{\rm 1st},T),
\label{eq.1st}
\end{equation}
where ${\bar \Delta}$ satisfies the stationary condition in Eq. (\ref{stationaryA}) (solid circles in Fig. \ref{fig2}(b2)). In this case, the superfluid order parameter {\it discontinuously} becomes non-zero at $h=h_{\rm c}^{\rm 1st}$, as schematically shown in Fig. \ref{fig2}(c2). 
\par
In principle, the FF state with ${\bm Q}\ne 0$ may be possible as the superfluid state in the right hand side of Eq. (\ref{eq.1st}). However, within our numerical results, the BCS state always satisfies Eq. (\ref{eq.1st}) before the single-$\bm{Q}$ FF state (that is, $\Omega_{\rm MF}({\bar \Delta}, \bm{Q}=0)<\Omega_{\rm MF}({\bar \Delta}, \bm{Q}\neq 0)$ is always satisfied) \cite{note.crystal2}. Regarding this, we will later explain that Eq. (\ref{eq.1st}) also determines the phase boundary between the normal state and the phase separation into the normal and the superfluid states. In this sense, our numerical results are consistent with the experimental fact that the phase separation into the BCS state and the polarized normal state is always observed in spin-polarized Fermi gases \cite{Shin2006,Zwierlein2006,Partridge2006,Shin2008,Mitra2016}. (If the FF state satisfies Eq. (\ref{eq.1st}) before the BCS state, a phase separation into the {\it FF state} and the polarized normal state would occur.) Thus, in this paper, we only consider the BCS case in Eq. (\ref{eq.1st}).
\par
\par
\subsubsection{Phase separation in $T$-$P$ phase diagram \label{Sec.TP.Th}}
\par
The $T$-$h$ phase diagram is useful for the study of the FFLO state in metallic superconductivity, because the misalignment of the two Fermi surfaces is tuned by an external magnetic field $h$. On the other hand, $T$-$P$ phase diagram is convenient in the Fermi gas case, because the spin imbalance in this case is experimentally tuned by directly adjusting the polarization $P$ \cite{Shin2006,Zwierlein2006,Partridge2006,Shin2008,Mitra2016}.
\par
Regarding the $T$-$P$ phase diagram, we point out that the coupled equations (\ref{eq.2nd_gap}) with (\ref{eq.num_normal}) can still be used to determine the second-order phase transition line in this phase diagram. That is, once the critical magnetic field $h_{\rm c}^{\rm 2nd}$ in the $T$-$h$ phase diagram is obtained by solving these equations, the corresponding critical polarization $P_{\rm c}$ in the $T$-$P$ phase diagram can be immediately obtained by evaluating Eq. (\ref{eq.polarization}) at $h_{\rm c}^{\rm 2nd}$.
\par
On the other hand, we need to be careful in dealing with the first-order phase transition: At $h_{\rm c}^{\rm 1st}$ in the $T$-$h$ phase diagram in Fig. \ref{fig2}(a), we should note that the number $N$ of Fermi atoms at $\Delta=0$ is usually different from that at $\Delta={\bar \Delta}$, for a fixed value of the averaged chemical potential $\mu$. This means that, when $N$ is fixed as in the cold Fermi gas experiments \cite{Shin2006,Zwierlein2006,Partridge2006,Shin2008,Mitra2016}, we have {\it two} different values of $\mu$ at the first-order phase transition: (1) $\mu_{\rm N}$ that is obtained by approaching this phase boundary from the normal state (${\bar \Delta}=0$). (2) $\mu_{\rm SF}$ that is obtained by approaching this phase boundary from the superfluid state (${\bar \Delta}\ne 0$). These give different values of critical polarizations, $P_{\rm c}(N,\mu_{\rm N},T)$ and $P_{\rm c}(N,\mu_{\rm SF},T)$ ($<P_{\rm c}(N,\mu_{\rm N},T)$). In the region between the two, $P_{\rm c}(N,\mu_{\rm SF},T)<P<P_{\rm c}(N,\mu_{\rm N},T)$, neither the normal state nor the superfluid state can satisfy the constraint on the fixed $N$, indicating the occurrence of the phase separation into the superfluid state and the (polarized) normal state \cite{Sheehy2006,He2008}. In other words, the first-order phase boundary in the $T$-$h$ phase diagram in Fig. \ref{fig2}(a) splits into two phase boundaries in the $T$-$P$ phase diagram \cite{Sheehy2006,He2008}: One is the boundary between the normal state and the phase separation, and the other is the boundary between the superfluid state and the phase separation.
\par
For the phase separation, this paper only deals with the boundary between the phase separation and the normal state. In this case, this boundary can be determined by solving the number equation (\ref{eq.num_normal}), together with Eq. (\ref{eq.1st}) \cite{note.Pc1}. As mentioned previously, the phase separation in our case is always into the {\it BCS state} and the spin-polarized normal state, as observed experimentally \cite{Shin2006,Zwierlein2006,Partridge2006,Shin2008,Mitra2016}.
\par
\par
\subsubsection{Computational Remarks}
\par
Before ending this section, we comment on the parameter region that we consider in this paper: (1) For the filling fraction $n$, we restrict our computations to the low filling case ($n<1$). This is because, although the nested Fermi surface near the half-filling $(n=1)$ strongly enhances fluctuations in the particle-hole channel when $t'=0$ \cite{Tamaki2008}, the present HNSR theory only takes into account fluctuations in the Cooper channel. (2) For the interaction strength $U$, we only deal with the weak-coupling case, because the NSR theory is known to unphysically give negative spin susceptibility \cite{Pantel2016, Kashimura2012}, as well as negative polarization for $h=[\mu_\up-\mu_\down]/2>0$, in the strong-coupling regime. (We have numerically confirmed that the same problems also exist in the HNSR theory.) However, since the FFLO state is expected in the weak-coupling regime where large spin-$\up$ and spin-$\down$ Fermi surfaces exist, the HNSR theory is still applicable to the study of the FFLO state, even under this restriction. (3) Because of computational problems, the temperature region is restricted down to $T/T_{\rm c}\simeq 0.1$.
\par
\begin{figure*}[tb]
\centering
\includegraphics[width=15cm]{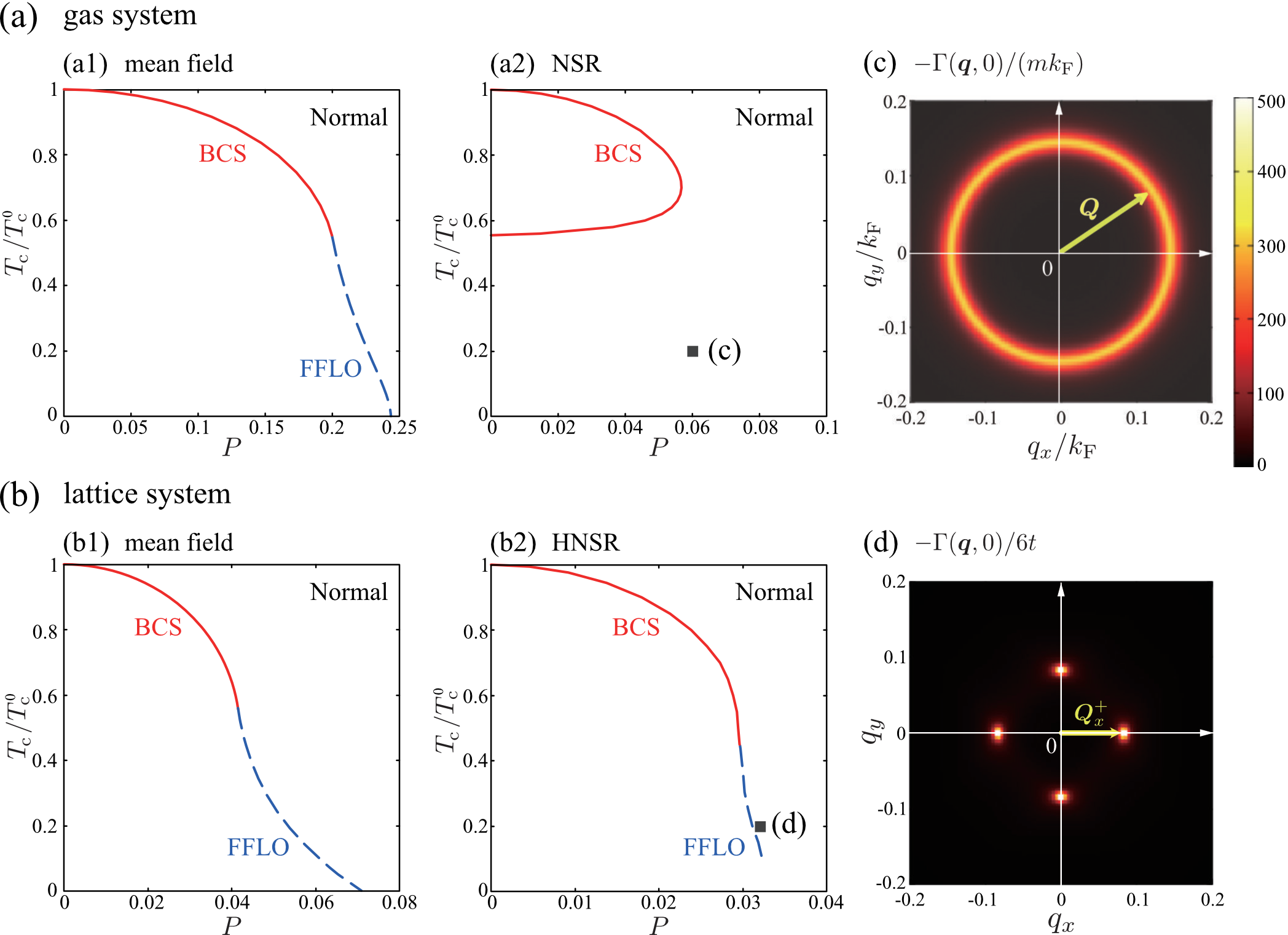}
\caption{Calculated second-order superfluid phase transition temperature $T_{\rm c}$. In this figure, we fix the total number $N=N_\up+N_\down$ of Fermi atoms. $T_{\rm c}$ is normalized by the value at $P=0$ ($\equiv T_{\rm c}^0$). (a) Fermi gas in the absence of optical lattice, when $(k_{\rm F}a_s)^{-1}=-1$. (Here, $a_s$ and $k_{\rm F}$ are the $s$-wave scattering length and the Fermi momentum, respectively.) (b) Lattice Fermi gas, when $U/(6t)=0.4$, $n=0.3$, and $t'=0$. Panels (a1) and (b1) show the mean-field results (where pairing fluctuations are ignored). Panels (a2) and (b2) show the results including pairing fluctuations within the NSR and HNSR theories, respectively. The solid (dashed) line is the phase boundary between the BCS (FFLO) state and the normal state. Panels (c) and (d) show the intensity of the particle-particle scattering matrix $\Gamma({\bm q},i\nu_n=0)$ in Eq. (\ref{eq.gamma}) at the positions (c) and (d) shown in panels (a2) and (b2), respectively. In panel (d), the lattice constant is taken to be unity (same for Fig. \ref{fig4}, Fig. \ref{fig5} and Fig. \ref{fig6}). The phase separation is ignored in this figure.
}
\label{fig3}
\end{figure*}
\par
\section{Feasibility of FFLO state in spin-imbalanced lattice Fermi gas}
\par
In Sec. III.A, we first examine how the optical lattice stabilizes the FFLO state, within the ignorance of the phase separation. The competition between this stabilized FFLO state and the phase separation is discussed in Sec. III.B. 
\par
\subsection{Stabilization of FFLO state against FFLO pairing fluctuations}
\par
Figure \ref{fig3} shows the second-order superfluid phase transition temperature $T_{\rm c}$ in a spin-balanced Fermi gas in the weak-coupling regime \cite{note.U}. In the absence of the cubic optical lattice, the FFLO superfluid phase transition obtained in the mean-field theory (panel (a1)) vanishes, when one includes effects of pairing fluctuations, as shown in panel (a2). (We summarize the NSR theory to obtain panel (a2) in Appendix A.) This vanishing FFLO state is consistent with the previous work, stating the complete destruction of the FFLO long-range order by strong FFLO pairing fluctuations \cite{Shimahara1998,Ohashi2002,Radzihovsky2009,Radzihovsky2011,Yin2014,Jakubczyk2017,Wang2018,Zdybel2021}. 
\par
Figures \ref{fig3}(b1) and (b2) show the revival of the FFLO state, when the system is loaded on the cubic optical lattice. However, the critical polarization $P_{\rm c}$ at the superfluid instability in Fig. \ref{fig3}(b2) is overall smaller than the mean-field result in Fig. \ref{fig3}(b1), indicating that FFLO pairing fluctuations still disturb the FFLO phase transition to some extent there.
\par
To explain the background physics of the vanishment and revival of the FFLO state seen in Figs. \ref{fig3}(a2) and (b2), we recall that, when Eq. (\ref{eq.2nd_gap}) is satisfied, the particle-particle scattering matrix $\Gamma({\bm q}={\bm Q},i\nu_m=0)$ in Eq. (\ref{eq.gamma}) always diverges. In the absence of optical lattice, because of the spacial isotropy of the system, $\Gamma({\bm q},i\nu_m)$ behaves as, around $({\bm q},\nu_m)=({\bm Q},0)$ \cite{Ohashi2002},
\begin{equation}
\Gamma({\bm q},i\nu_m) \simeq  
\frac{-U}{\gamma \big[|{\bm q}|-|{\bm Q}|\big]^2 +\lambda|\nu_m|},  
\end{equation}
where $\gamma = -(U/2)\partial^2\Pi({\bm q}, 0)/\partial q^2|_{{\bm q}\to{\bm Q}}$ and $\lambda= -U\partial{\rm Im}\Pi({\bm Q},i\nu_n\to z+i\delta)/\partial z|_{z\to 0}$. Substituting this into the NSR number equation (\ref{eq.gas.NSR.n}), one finds
\begin{align}
N_\sigma
&\simeq
\sum_{\bm k}f(\xi_{{\bm k},\sigma})
+
T \frac{\partial \Pi({\bm Q}, 0)}{\partial \mu_\sigma}
\notag\\
&\hspace{0.5cm}\times
\sum_{\bm{q},i\nu_m} e^{i\nu_m \delta}
\frac{U}{\gamma \big[|{\bm q}|-|{\bm Q}|\big]^2 +\lambda|\nu_m|}
\nonumber\\[6pt]
&\simeq
\sum_{\bm k}f(\xi_{{\bm k},\sigma})
+
{UT \over \gamma}
\frac{\partial \Pi({\bm Q}, 0)}{\partial \mu_\sigma}
\int_0^{q_{\rm c}}{q^2dq \over 2\pi^2}\frac{1}{[|{\bm q}|-|{\bm Q}|\big]^2},
\label{eq.gas.nFL.app2}
\end{align}
where $q_{\rm c}$ is a cutoff, and we have only retained the most singular term with $\nu_m=0$ in obtaining the last line, for simplicity. The last term in Eq. (\ref{eq.gas.nFL.app2}) always diverges unless ${\bm Q}=0$, which means that Eqs. (\ref{eq.2nd_gap}) and (\ref{eq.gas.NSR.n}) are never satisfied simultaneously. Since $\Gamma({\bm Q},i\nu_m)$ physically describes pairing fluctuations with the center-of-mass momentum ${\bm Q}$, this result is interpreted as the destruction of the FFLO long-range order by FFLO pairing fluctuations. Indeed, at the position (c) in Fig. \ref{fig3}(a2), Fig. \ref{fig3}(c) shows that $\Gamma({\bm q},0)$ has large intensity around $|{\bm q}|=|{\bm Q}|$, indicating the existence of strong FFLO pairing fluctuations there. The ring structure seen in this figure reflects the degeneracy of FFLO pairing fluctuations in terms of the direction of ${\bm Q}$ in the absence of optical lattice.
\par
The divergence of $\Gamma({\bm Q},0)$ also occurs in the lattice Fermi gas, when Eq. (\ref{eq.2nd_gap}) is satisfied; however, the number of ${\bm Q}$-vectors that satisfy Eq. (\ref{eq.2nd_gap}) becomes {\it finite} in this case, reflecting the discrete four-fold rotational symmetry of the cubic lattice system around the $x$-, $y$-, and $z$-axis. Indeed, at the position (d) shown in Fig. \ref{fig3}(b2), Fig. \ref{fig3}(d) shows that the intensity of $\Gamma({\bm q},0)$ directly reflects this symmetry property. Denoting the ${\bm Q}$-vectors that satisfy Eq. (\ref{eq.2nd_gap}) as $\bm{Q}_{\alpha=x,y,z}^{\pm}$ ($\parallel \pm \bm{i}_{\alpha}$, where $\bm{i}_\alpha$ is the primitive vector along the $\alpha$-axis), one can approximate $\Gamma({\bm q}\simeq{\bm Q}_\alpha^\pm ,i\nu_m)$ to
\begin{equation}
\Gamma(\bm{q},i\nu_m) \simeq  \frac{-U}{\tilde{\gamma} [\bm{q} -\bm{Q}^\pm_\alpha]^2 +\tilde{\lambda}|\nu_m|},
\label{eq.gamma2}
\end{equation}
where $\tilde{\gamma} =-(U/2)\nabla_{\bm{q}}^2\Pi(\bm{q}, 0)|_{{\bm q}\to{\bm Q}^\pm_\alpha}$ and $\tilde{\lambda}= -U\partial {\rm Im}\Pi(\bm{Q}_\alpha^\zeta,i\nu_m\to z+i\delta)/\partial z|_{z\to 0}$. Substitution of Eq. (\ref{eq.gamma2}) into the number equation (\ref{eq.num_normal}) gives
\begin{align}
N_\sigma
\simeq &
\sum_{\bm k}f(\xi_{{\bm k},\sigma})+
{TU \over \tilde{\gamma}} \sum_{\eta=\pm,\alpha=x,y,z,\sigma'}
\left(
{\partial \Pi(\bm{Q}_\alpha^\eta, 0) \over \partial {\tilde \mu}_{\sigma'}}
\right)
\notag\\
&\times
\left(
{\partial {\tilde \mu}_{\sigma'} \over \partial\mu_\sigma}
\right)
\sum_{\bm q}^{q_{\rm c}}{1 \over [\bm{q}-\bm{Q}_\alpha^\eta]^2}
\nonumber
\\
&-
TU\sum_{{\bm q},i\nu_m,\sigma'}
e^{i\delta\nu_m}
\left(
{\partial \Pi({\bm q},i\nu_m) \over \partial{\tilde \mu}_{\sigma'}}
\right)
\left(
{\partial {\tilde \mu}_{\sigma'} \over \partial\mu_\sigma}
\right).
\label{eq.tilde.n.app}
\end{align}
Replacing ${\bm q}-{\bm Q}_\alpha^\eta$ by ${\bm q}$ in the second term in Eq. (\ref{eq.tilde.n.app}), we find that this term converges in three dimension.  Since the last term in Eq. (\ref{eq.tilde.n.app}) also converges, the coupled equations (\ref{eq.2nd_gap}) and (\ref{eq.num_normal}) can have simultaneous solutions for the FFLO state in three dimension, when the system in loaded on the cubic optical lattice \cite{noteMermin}.
\par
We note that, while the six solutions specified by ${\bm Q}_{\alpha=x,y,z}^\pm$-vectors are all degenerate at the second-order FFLO phase transition, this degeneracy is lifted in the FFLO phase. For the FF state \cite{Fulde1964}, one of ${\bm Q}_\alpha^\pm$ is chosen (for example, $\bm{Q}_x^+$). For a FFLO state \cite{Larkin1964}, a pair of them, e.g., $(\bm{Q}_x^+,{\bm Q}_x^-)$, is chosen. To identify what combination is realized, we need to evaluate the free energy of each candidate in the superfluid state below $T_{\rm c}$, which remains as one of our future problems.
\par
\begin{figure}[t]
\centering
\includegraphics[width=8.3cm]{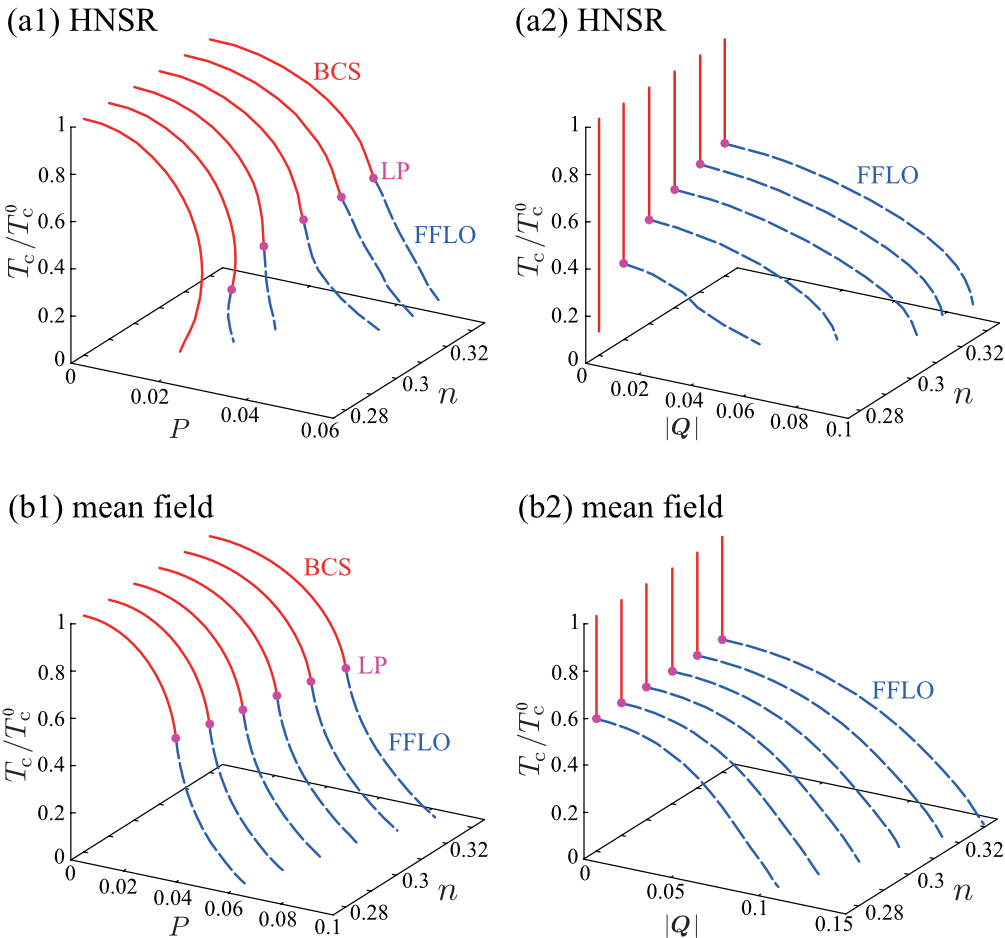}
\caption{Second-order superfluid phase transition temperature $T_{\rm c}$ in a lattice Fermi gas, as a function of the polarization $P$ and the filling fraction $n$. (a1) HNSR theory. (b1) Mean-field BCS theory. The solid (dashed) line shows the BCS (FFLO) phase transition. The solid circle shows the Lifshitz point (LP), at which three kinds of phase boundaries between (1) the BCS and the normal states, (2) the FFLO and the normal states, as well as (3) the BCS and the FFLO states, meet one another. Panels (a2) and (b2) show the magnitude of the ${\bm Q}$-vector along the $T_{\rm c}$-line shown in panels (a1) and (b2), respectively. We set $U/(6t)=0.4$ and $t'=0$. $T_{\rm c}^0$ is the superfluid phase transition temperature when $P=0$. 
}
\label{fig4}
\end{figure}
\par
Figure \ref{fig4} shows how the filling fraction $n$ affects the second-order superfluid phase transition temperature $T_{\rm c}$ in the presence of optical lattice. In the low filling region ($n\lesssim 0.3$), comparing the upper HNSR results with the lower mean-field ones in this figure, we find that the FFLO state is still remarkably suppressed by FFLO pairing fluctuations, even in the presence of optical lattice. In the mean-field theory, the Lifshitz point (solid circle in Fig. \ref{fig4}), at which the BCS phase transition changes to the FFLO one \cite{Chaikinbook,Hornreich1975}, is known to always appear at $T/T_{\rm c}^0\simeq 0.56$, regardless of the value of $n$ \cite{Machida1984, Shimahara1994, Fujita1984, Machida2006, Mizushima2007, Mizushima2014, Takahashi2014} (where $T_{\rm c}^0$ is the superfluid phase transition temperature at $P=0$). While this property is really seen in the lower mean-field results in Fig. \ref{fig4}, the upper panels show that it no longer holds, especially in the low filling regime, which also supports the existence of strong FFLO pairing fluctuations there.
\par
When $n\gesim 0.3$, although the critical polarization $P_{\rm c}$ at the FFLO phase transition (Fig. \ref{fig4}(a1)), as well as the magnitude of the FFLO ${\bm Q}$-vector at $T_{\rm c}$ (Fig. \ref{fig4}(a2)), are still smaller than the mean-field values, their qualitative behavior is found to be similar to the mean-field case shown in the lower panels in Fig. \ref{fig4}. This filling dependence implies that FFLO pairing fluctuations are more suppressed for higher filling cases. 
\par
\begin{figure}[tb]
\centering
\includegraphics[width=8.3cm]{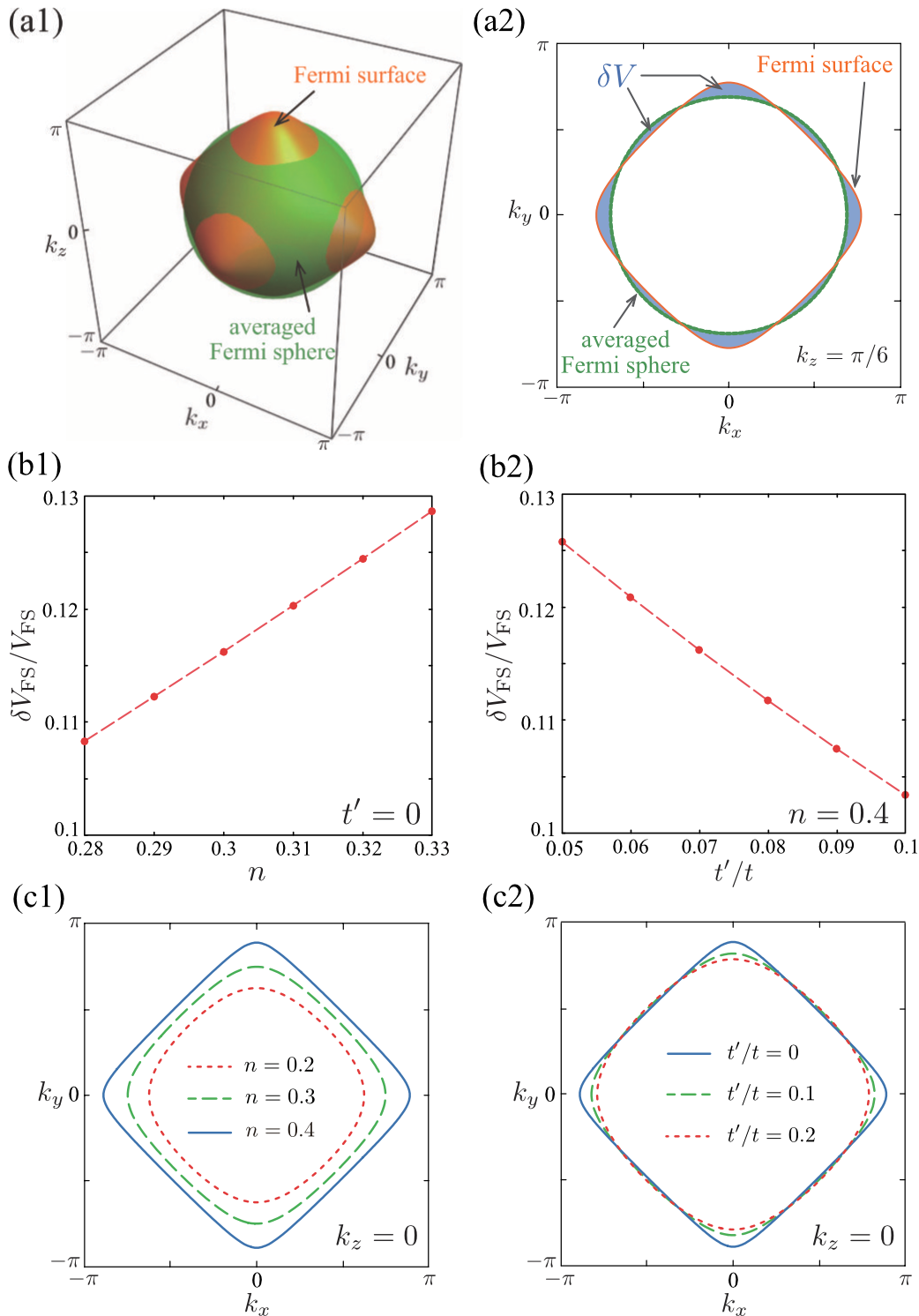}
\caption{Assessment of the anisotropy of Fermi surface. We introduce an averaged Fermi sphere that has the same volume $V_{\rm FS}$ as the anisotropic Fermi surface in a optical lattice (panel (a1)). We then sum up the {\it magnitude} $\delta V_{\rm FS}$ of the difference between these two Fermi surfaces at each momentum direction (panel (a2)). (b1) Calculated $\delta V_{\rm FS}/V_{\rm FS}$, as a function of the filling fraction $n$. We set $t'=0$. (b2) $\delta V_{\rm FS}/V_{\rm FS}$ as a function of the NNN hopping parameter $t'$, when $n=0.4$. For clarify, $n$-dependence and $t'$-dependence of the Fermi surface shape at $k_z=0$ are explicitly shown in panels (c1) and (c2), respectively.
}
\label{fig5}
\end{figure}
\par
We point out that the key to understand the above-mentioned filling dependence of the FFLO phase transition is the {\it anisotropy} of  Fermi surface. For example, in the low filling limit ($n\ll 1$), even in a lattice Fermi gas, particles near the Fermi level feel an almost {\it isotropic} Fermi surface, because the kinetic energy $\varepsilon_{\bm k}$ in Eq. (\ref{eq.band}) with $t'=0$ behaves as, around ${\bm k}=0$,
\begin{equation}
\varepsilon_{{\bm k}\simeq 0}\simeq -6t+t{\bm k}^2.
\label{eq.Fermi0}
\end{equation}
As a result, strong FFLO pairing fluctuations still exist as in the absence of optical lattice, that remarkably disturb the FFLO phase transition there. 
\par
As one increases the filling fraction $n$, the Fermi surface shape gradually deviates from spherical, reflecting the detailed momentum dependence of the band dispersion $\varepsilon_{\bm k}$ in Eq. (\ref{eq.band}). To simply quantify this deformation of Fermi surface, we introduce an averaged Fermi sphere that has the same volume $(\equiv V_{\rm FS}$) as the Fermi surface of the system, to measure the magnitude $\delta V_{\rm FS}$ of the difference between these two Fermi surfaces in all momentum directions (see Figs. \ref{fig5}(a1) and (a2)). Then, Fig. \ref{fig5}(b1) shows that the increase of the filling fraction done in Fig. \ref{fig4} really enhances the anisotropy $\delta V_{\rm FS}/V_{\rm FS}$ of the Fermi surface (see also Fig. \ref{fig5}(c1)). This enhancement weakens FFLO pairing fluctuations, to more stabilize the FFLO state, which explains the filling dependence of the FFLO phase transition temperature seen in Fig. \ref{fig4}(a1). 
\par
\begin{figure}[t]
\centering
\includegraphics[width=8.3cm]{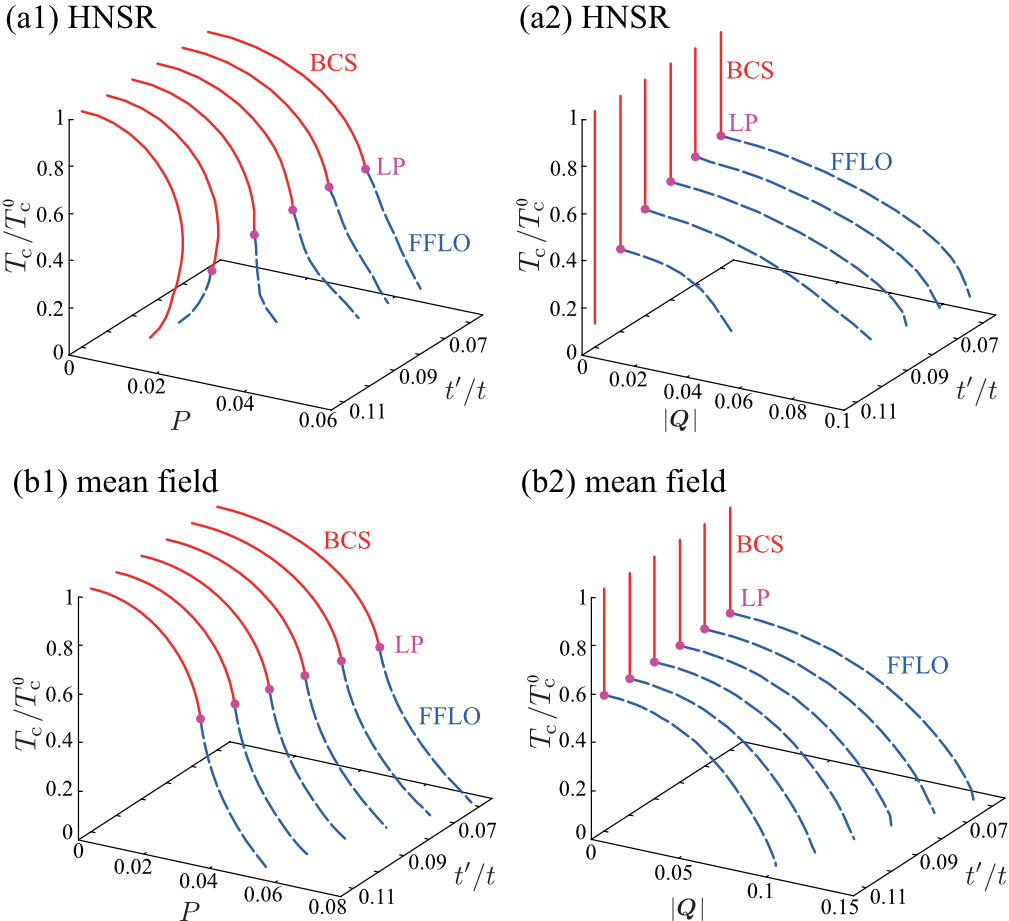}
\caption{Same plots as Fig. \ref{fig4} for various values of the NNN hopping amplitude $t'$, when $n=0.4$.}
\label{fig6}
\end{figure}
\par
To support the above explanation, it is useful to consider effects of the NNN hopping $t'$. As shown in Fig. \ref{fig6}(a1) and (a2), the increase of $t'$ {\it suppresses} the FFLO phase transition. Regarding this result, we note that, while the increase of the filling fluctuation $n$ enhances the anisotropy of the Fermi surface, the increase of $t'$ tends to round the Fermi surface, as shown in Figs. \ref{fig5}(b2) and (c2). In addition, the mean-field results shown in the lower panels in Fig. \ref{fig6} only weakly depend on $t'$. Thus, the suppression of the FFLO phase transition with increasing $t'$ can also be explained as a result of the shape change of the Fermi surface by the NNN hoping.
\par
In the current stage of cold Fermi gas physics, a superfluid $^6$Li Fermi gas in an optical lattice has been realized, only when the lattice potential is very shallow \cite{Ketterle_lattice}. Regarding this, although this experimental situation is somehow different from the simple Hubbard model in Eq. (\ref{eq.H}), our results indicate that the crucial key to stabilize the FFLO state is, not the detailed lattice potential, but the resulting anisotropy of Fermi surface. Thus, if such a shallow optical lattice can still deform the Fermi surface enough to suppress the FFLO pairing fluctuations,  the FFLO superfluid Fermi gas would be realized there.
\par
We note that the importance of Fermi surface for the stability of the FFLO state has so far mainly been discussed in terms of the nesting effect \cite{Shimahara1994, Matsuda2007, Shimahara2021}: A nested Fermi surface is preferable for the stabilization of the FFLO state, being analogous to the nesting effects on the charge-density-wave, as well as the spin-density-wave states. Besides this, our results suggest that the anisotropy of Fermi surface is also important to stabilize the FFLO state against FFLO pairing fluctuations.
\par
\begin{figure*}[tb]
\centering
\includegraphics[width=15.5cm]{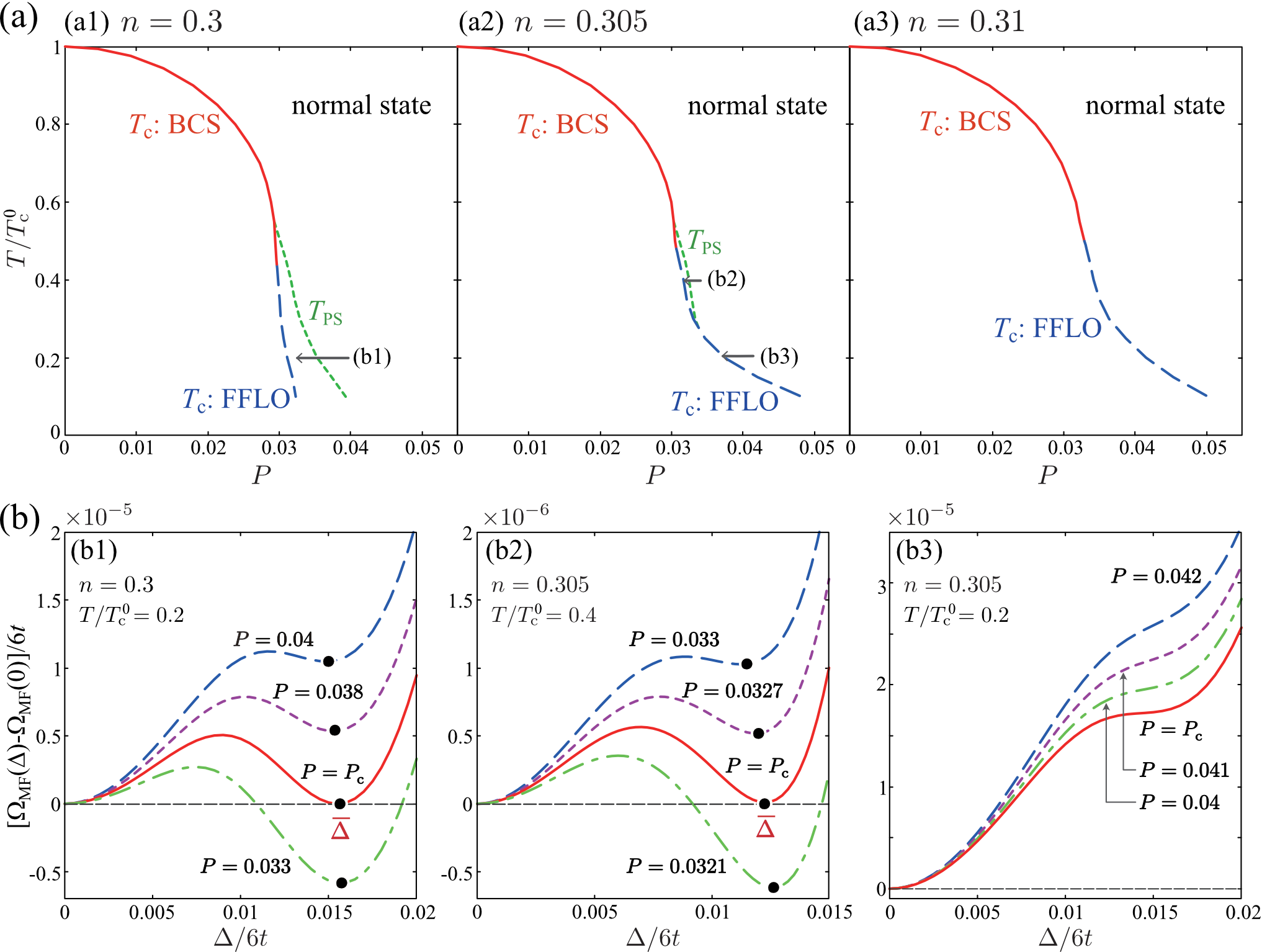}
\caption{(a) HNSR results on the second-order superfluid phase transition temperature $T_{\rm c}$ (solid line: BCS transition. dashed line: FFLO transition), as well as the phase separation temperature $T_{\rm PS}$ (dotted line). In each upper panel, the filling fraction $n$ is fixed ($n=$0.3, 0.305, and 0.31). Below $T_{\rm PS}$, the system separates into the BCS and the spin-polarized normal states. (b) Thermodynamic potential $\Omega_{\rm MF}(\Delta)-\Omega_{\rm MF}(0)$, measured from the value at $\Delta=0$. Each panel shows the $\Delta$-dependence along the path (b1)-(b3) shown in the upper panels. In panels (b1) and (b2), ${\bar \Delta}~(>0)$ is the position at which $\Omega_{\rm MF}(\Delta)$ takes a local minimum (solid circle), and $P_{\rm c}$ is the polarization at the phase separation line. $P_{\rm c}$ in panel (b3) denotes the polarization at the FFLO superfluid phase transition. We set $U/(6t)=0.4$ and $t'=0$.
}
\label{fig7}
\end{figure*}
\par
\begin{figure}[t]
\centering
\includegraphics[width=7.8cm]{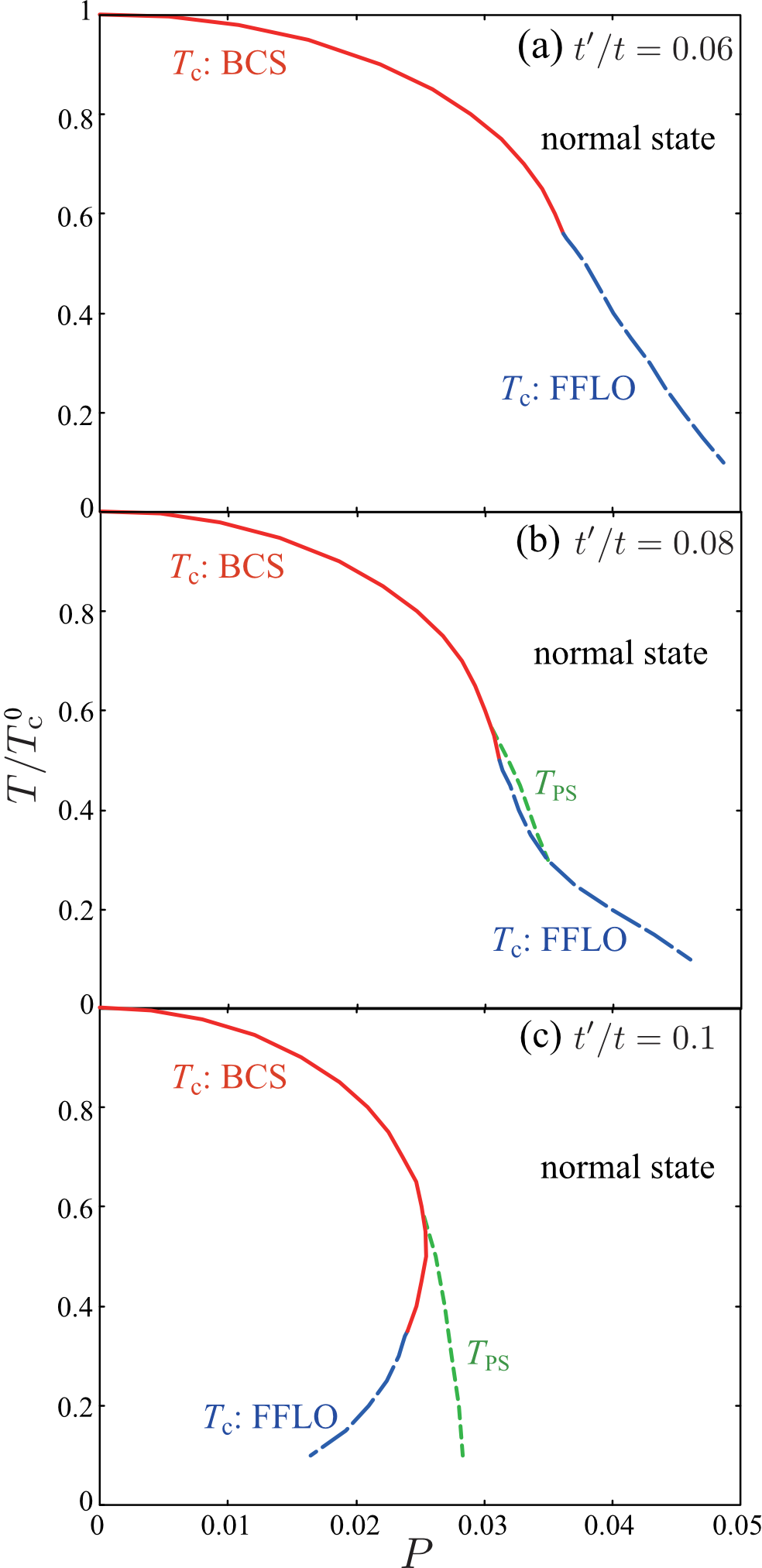}
\caption{Same plots as Figs \ref{fig7}(a1)-(a3) for different values of NNN hopping $t'/t$, when $n=0.4$.
}
\label{fig8}
\end{figure}
\par
\begin{figure}[tb]
\centering
\includegraphics[width=7.8cm]{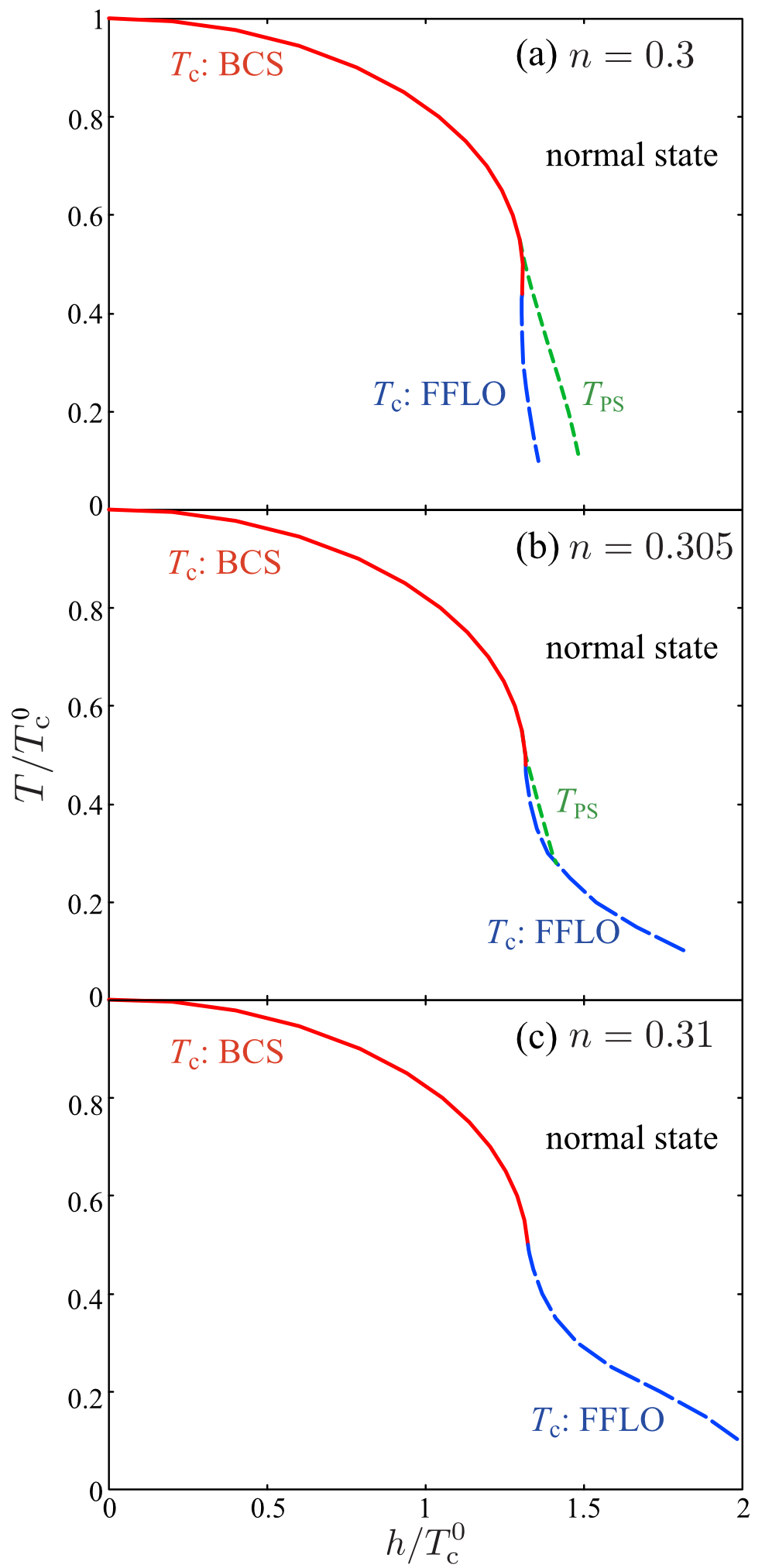}
\caption{
Same plots as Fig. \ref{fig7}(a) in the $T$-$h$ phase diagram. Since the filling fraction $n$ is fixed in each panel, the phase separation occurs at $T_{\rm PS}$ as in the $T$-$P$ phase diagram in Fig. \ref{fig7}(a). Regarding this, we note that, when $h$ is decreased from the phase boundary at $T_{\rm PS}$ under the condition that the averaged chemical potential $\mu$ is fixed, the BCS first-order phase transition occurs, without the phase separation.}
\label{fig9}
\end{figure}
\par
\subsection{Competition between FFLO state and phase separation}
\par
We next examine whether or not the stabilized FFLO state can also overcome the competition with the phase separation in a spin-imbalance Fermi gas. When $(U/6t,n,t')=(0.4,0.3,0)$, Fig. \ref{fig7}(a1) shows that the $T_{\rm c}$-line of the FFLO state is always located on the inside of the phase-separation temperature $T_{\rm SP}$. [$\Omega_{\rm MF}(0)=\Omega_{\rm MF}(\bar{\Delta})$ is satisfied at $T_{\rm PS}$ ($P=P_{\rm c}$) as shown in Fig. \ref{fig7}(b1).] That is, starting from the normal state, the system  always experiences the phase separation, before reaching the FFLO superfluid phase transition. Regarding this result, we note that $T_{\rm c}$ in this figure is determined within the ignorance of the phase separation. Thus, although the $T_{\rm c}$-line of the FFLO state is still drawn on the left side of the $T_{\rm PS}$-line in Fig. \ref{fig7}(a1), it does {\it not} mean the occurrence of the FFLO phase transition, when one further decreases the polarization $P$ from the phase separation line.
\par
When the anisotropy of the Fermi surface is enhanced by increasing the filling fraction $n$, the unwanted FFLO pairing fluctuations are suppressed, so that the phase boundary of the FFLO state shifts toward the higher polarization regime. On the other hand, since the phase separation phenomenon is dominated by the spin imbalance of the system, $T_{\rm PS}$ is not so sensitive to $n$, compared to the FFLO phase transition. Thus, the $T_{\rm c}$-line of the FFLO state eventually exceeds the phase-separation line, as shown in Figs. \ref{fig7}(a2) and (a3). When $n=0.305$ (panel (a2)), while the phase separation still masks the FFLO state in the temperature region, $0.28 \lesssim T/T^0_{\rm c} \lesssim 0.55$, one can reach the FFLO superfluid phase transition, when $T/T^0_{\rm c} \lesssim 0.28$. [See also the difference between the Fig. \ref{fig7}(b2) and (b3).] In the case of $n=0.31$ shown in panel (a3), we can always reach the FFLO state when $T/T_{\rm c}^0\lesssim 0.5$, without being disturbed by the phase separation. 
\par
Figure \ref{fig8} shows how the NNN hopping $t'$ affects the competition between the FFLO state and the phase separation. As expected from the property that $t'$ tends to round the Fermi surface shape (see Figs. \ref{fig5}(b2) and (c2)), the phase boundary of the FFLO state shifts toward the left, as one moves from panel (a) to (c). As a result, while the FFLO phase transition is possible when $t'/t=0.06$, it is completely masked by the phase separation when $t'/t=0.1$, due to the insufficient suppression of FFLO pairing fluctuations by a nearly spherical Fermi surface.
\par
The recovery of the FFLO state by the enhancement of the anisotropy of the Fermi surface can also be seen in the $T$-$h$ phase diagram as shown in Fig. \ref{fig9}. We note that, as in the $T$-$P$ phase diagram shown in Figs. \ref{fig7}(a1) and (a2), the phase separation occurs at $T_{\rm PS}$ in panels (a) and (b), when the filling fraction $n$ is fixed \cite{He2008}. On the other hand, when one decreases $h$ from the phase boundary at $T_{\rm ps}$ under the condition that the average chemical potential $\mu$ is fixed, such phase separation does not occur. Instead, the system experiences the first-order phase transition into the BCS state at $T_{\rm PS}$, as discussed in Sec. \ref{Sec.TP.Th}.
\par
\section{Summary}
\par
To summarize, we have discussed the feasibility of the FFLO state in a spin-imbalanced Fermi gas loaded on a cubic optical lattice. The current cold Fermi gas physics is facing two difficulties in achieving the FFLO state: (1) Desperate destruction of FFLO long-range order by FFLO pairing fluctuations. (2) Competition between the FFLO state and the phase separation. In this paper, we theoretically explore a possible route to realize an FFLO superfluid Fermi atomic gas, overcoming these difficulties. For this purpose, going beyond the mean-field theory, we included FFLO pairing fluctuations, by extending the strong-coupling theory developed by Nozi\'{e}res-Schmitt-Rink (NSR) to a three-dimensional spin-imbalanced attractive Hubbard model. 
\par
We showed that the FFLO state becomes stable, when the Fermi gas is loaded on a cubic optical lattice; however, the FFLO phase transition is still sensitive to the filling fraction $n$, especially in the low-filling regime. To understand this, we pointed out the importance of the anisotropy of the Fermi surface: When the Fermi surface is deformed from the sphere by varying the filling fraction and the next-nearest-neighbor hopping amplitude, the unwanted FFLO pairing fluctuations become weak, which promotes the stabilization of the FFLO long-range order. 
\par
For this stabilized FFLO state, we have further examined whether or not it can still survive, when the possibility of the phase separation is taken into account. We clarified that, starting from the normal state, we can reach the FFLO phase transition without being disturbed by the phase separation, when the anisotropy of the Fermi surface remarkably suppresses FFLO pairing fluctuations.
\par
In this paper, we have only examined the phase boundary of the FFLO state. Extension of the present theory to the FFLO superfluid state below $T_{\rm c}$ is an interesting future challenge. Even for the study of the phase boundary, we need to extend to the present strong-coupling scheme to also include fluctuations in the particle-hole channel, in order to access the region near the half-filling  (where the nested Fermi surface is known to enhance charge-density-wave fluctuations \cite{Tamaki2008}). We also note that the possibility of the FFLO state has been discussed in unitary Fermi gases \cite{Son2006, Bulgac2008, Tuzemen2020}. Based on a non-perturbative approach, it has been shown that a splitting point (where the uniform superfluid, a gapless superfluid, and the FFLO superfluid phases meet) exists in the BEC regime, which indicates a robust FFLO phase around the unitary limit \cite{Son2006}. However, it is known that the NSR theory is not applicable to the unitary regime in the presence of spin imbalance, because it unphysically gives negative spin susceptibility when $(k_{\rm F}a_s)\gesim -0.5$ \cite{Kashimura2012}. Thus, to examine the FFLO state in this regime, we need a more sophisticated strong-coupling theory, such as the self-consistent $T$-matrix approximation. Although these future problems still remain, our results would provide a useful clue for cold Fermi gas experiments toward the achievement of the FFLO state. Since this unconventional Fermi superfluid is also discussed in condensed matter physics, as well as high-energy physics, the realization of a FFLO superfluid Fermi atomic gas would make a great impact on these research fields.
\par
\begin{acknowledgments}
We thank K. Manabe and K. Nishimura for stimulating discussions. T.K. was supported by MEXT and JSPS KAKENHI Grant-in-Aid for JSPS fellows (No.JP21J22452). Y.O. was supported by a Grant-in-aid for Scientific Research from MEXT and JSPS in Japan (No.JP18K11345, No.JP18H05406, No.JP19K03689, and JP22K03486).
\end{acknowledgments}
\par
\appendix
\section{NSR theory for a spin-imbalanced Fermi gas in the absence of optical lattice}
\par
We explain the outline of the NSR theory for a spin-imbalanced Fermi gas in the absence of optical lattice \cite{Liu2006,Parish2007}, which is used to obtain Fig. \ref{fig3}(a2). We start from the BCS Hamiltonian,
\begin{align}
H =& 
\sum_{{\bm k},\sigma=\up,\down} \xi_{\bm{k},\sigma} \hat{c}_{\bm{k},\sigma}^\dagger \hat{c}_{\bm{k},\sigma} 
\notag\\
&
-U\sum_{\bm{k}, \bm{k}', \bm{q}} \hat{c}_{\bm{k}+\bm{q}/2,\up}^\dagger \hat{c}_{-\bm{k}+\bm{q}/2,\down}^\dagger \hat{c}_{-\bm{k}'+\bm{q}/2, \down} \hat{c}_{\bm{k}'+\bm{q}/2, \up}.
\label{app.H}
\end{align}
Here, $\hat{c}_{\bm{k},\sigma}$ is the annihilation operator of a Fermi atom with pseudo-spin $\sigma=\up, \down$. $\xi_{\bm{k},\sigma}=\bm{k}^2/2m -\mu_\sigma=\varepsilon_{\bm k}-\mu_\sigma$ is the kinetic energy, measured from the Fermi chemical potential $\mu_{\sigma}$ (where $m$ is an atomic mass). In the continuum case, it is convenient to measure the interaction strength $-U$ in terms of the $s$-wave scattering length $a_s$, which is related to the pairing interaction $-U$ as \cite{Randeria}
\begin{equation}
\frac{4\pi a_s}{m} = \frac{-U}{1- U\sum_{\bm{k}}^{k_{\rm c}} \frac{1}{2\ep_{\bm{k}}}},
\label{eq.as}
\end{equation}
where $k_{\rm c}$ is a momentum cutoff. In Fig. \ref{fig3}(a2), we take $(k_{\rm F}a_s)^{-1}=-1$, which corresponds to the weak-coupling BCS regime (where $k_{\rm F}=(3\pi^2 N)^{1/3}$ is the Fermi momentum of a spin-balanced two-component Fermi gas with $N$ atoms).
\par
In the NSR theory, the $T_{\rm c}$-equation is given by Eq. (\ref{eq.2nd_gap}) where the Hartree-shifted kinetic energy ${\tilde \xi}_{{\bm k},\sigma}$ involved in $\Pi$ in Eq. (\ref{eq.Pi.N}) is replaced by $\xi_{{\bm k},\sigma}$. Since this equation exhibits the ultraviolet divergence, we renormalize it by using the $s$-wave scattering length $a_s$ in Eq. (\ref{eq.as}) \cite{Randeria}.
\par
We solve the (renormalized) $T_{\rm c}$-equation, together with the NSR number equation,
\begin{align}
N_\sigma 
&=
\sum_{\bm{k}} f(\xi_{\bm{k},\sigma})  -T \sum_{\bm{Q}, i\nu_m} e^{i\nu_m \delta} 
\Gamma({\bm Q}, i\nu_m)\frac{\partial \Pi({\bm Q}, i\nu_m)}{\partial \mu_\sigma},
\label{eq.gas.NSR.n}
\end{align}
to self-consistently determine $T_{\rm c}$ and $\mu_\sigma$, for a given parameter set $((k_{\rm F}a_s)^{-1},N_\up,N_\down)$. In Eq. (\ref{eq.gas.NSR.n}), the particle-particle scattering matrix $\Gamma$ is given by Eq. (\ref{eq.gamma}) with ${\tilde \xi}_{{\bm k},\sigma}$ being replaced by $\xi_{{\bm k},\sigma}$. 
\par
\par


\begin{thebibliography}{99}
\bibitem{Fulde1964} P. Fulde and R. A. Ferrell, Phys. Rev. {\bf 135}, A550 (1964).
\bibitem{Larkin1964} A. I. Larkin and Y. N. Ovchinnikov, Zh. Eksp. Teor. Fiz. {\bf 47}, 1136 (1964) [Sov. Phys. JETP {\bf 20}, 762 (1965)].
\bibitem{Takada1969} S. Takada, and T. Izuyama, Prog. Theor. Phys. {\bf 41}, 635 (1969).
\bibitem{Takada1970} S. Takada, Prog. Theor. Phys. {\bf 43}, 27 (1970).
\bibitem{Shimahara1994} H. Shimahara, Phys. Rev. B {\bf 50}, 12760 (1994).
\bibitem{Bianchi2002} A. Bianchi, R. Movshovich, N. Oeschler, P. Gegenwart, F. Steglich, J. D. Thompson, P. G. Pagliuso, and J. L. Sarrao, Phys. Rev. Lett. {\bf 89}, 137002 (2002).
\bibitem{Bianchi2003} A. Bianchi, R. Movshovich, C. Capan, P. G. Pagliuso, and J. L. Sarrao, Phys. Rev. Lett. {\bf 91}, 187004 (2003).
\bibitem{Kumagai2006} K. Kumagai, M. Saitoh, T. Oyaizu, Y. Furukawa, S. Takashima, M. Nohara, H. Takagi, and Y. Matsuda, Phys. Rev. Lett. {\bf 97}, 227002 (2006).
\bibitem{Kitagawa2018} S. Kitagawa, G. Nakamine, K. Ishida, H. S. Jeevan, C. Geibel, and F. Steglich, Phys. Rev. Lett. {\bf 121}, 157004 (2018).
\bibitem{Singleton2000} J. Singleton, J. A. Symington, M.-S. Nam, A. Ardavan, M. Kurmoo, and P. Day, J. Phys. Condens. Matter {\bf 12}, L641 (2000).
\bibitem{Wright2011} J. A. Wright, E. Green, P. Kuhns, A. Reyes, J. Brooks, J. Schlueter, R. Kato, H. Yamamoto, M. Kobayashi, and S. E. Brown, Phys. Rev. Lett. {\bf 107}, 087002 (2011).
\bibitem{Mayaffre2014} H. Mayaffre, S. Kr\"{a}mer, M. Horvati\'{c}, C. Berthier, K. Miyagawa, K. Kanoda, and V. F. Mitrovi\'{c}, Nat. Phys. {\bf 10}, 928 (2014).
\bibitem{Cho2017} C.-w. Cho, J. H. Yang, N. F. Q. Yuan, J. Shen, T. Wolf, and R. Lortz, Phys. Rev. Lett. {\bf 119}, 217002 (2017).
\bibitem{Ok2020} J. M. Ok, C. I. Kwon, Y. Kohama, J. S. You, S. K. Park, J.-h. Kim, Y. J. Jo, E. S. Choi, K. Kindo, W. Kang, K. S. Kim, E. G. Moon, A. Gurevich, and J. S. Kim, Phys. Rev. B {\bf 101}, 224509 (2020).
\bibitem{Kasahara2020} S. Kasahara, Y. Sato, S. Licciardello, M. \v{C}ulo, S. Arsenijevi\'{c}, T. Ottenbros, T. Tominaga, J. B\"{o}ker, I. Eremin, T. Shibauchi, J. Wosnitza, N. E. Hussey, and Y. Matsuda, Phys. Rev. Lett. {\bf 124}, 107001 (2020).
\bibitem{Sheehy2007} D. E. Sheehy and L. Radzihovsky, Ann. Phys. {\bf 322}, 1790 (2007).
\bibitem{Kinnunen2018} J. J. Kinnunen, J. E. Baarsma, J.-P. Martikainen, and P. T\"{o}rm\"{a}, Rep. Prog. Phys. {\bf 81}, 046401 (2018).
\bibitem{Hu2006} H. Hu and X-J. Liu, Phys. Rev. A {\bf 73}, 051603(R) (2006).
\bibitem{Parish2007} M. M. Parish, F. M. Marchetti, A. Lamacraft, and B. D. Simons, Nat. Phys. {\bf 3}, 124 (2007).
\bibitem{Chevy2010} F. Chevy and C. Mora, Rep. Prog. Phys. {\bf 73}, 112401 (2010).
\bibitem{Vorontsov2007} A. B. Vorontsov and J. A. Sauls, Phys. Rev. Lett. {\bf 98}, 045301 (2007).
\bibitem{Aoyama2014} K. Aoyama, Phys. Rev. B {\bf 89}, 140502(R) (2014).
\bibitem{Wiman2016} J. J. Wiman and J. A. Sauls, J. Low Temp. Phys. {\bf 184}, 1054 (2016).
\bibitem{Levitin2019} L. V. Levitin, B. Yager, L. Sumner, B. Cowan, A. J. Casey, J. Saunders, N. Zhelev, R. G. Bennett, and J. M. Parpia, Phys. Rev. Lett. {\bf 122}, 085301 (2019).
\bibitem{Shook2020} A. J. Shook, V. Vadakkumbatt, P. Senarath Yapa, C. Doolin, R. Boyack, P. H. Kim, G. G. Popowich, F. Souris, H. Christani, J. Maciejko, and J. P. Davis, Phys. Rev. Lett. {\bf 124}, 015301 (2020).
\bibitem{Alford2001} M. Alford, J. A. Bowers, and K. Rajagopal, Phys. Rev. D {\bf 63}, 074016 (2001).
\bibitem{Casalbuoni2004} R. Casalbuoni and G. Nardulli, Rev. Mod. Phys. {\bf 76}, 263 (2004).
\bibitem{Anglani2014} R. Anglani, R. Casalbuoni, M. Ciminale, N. Ippolito, R. Gatto, M. Mannarelli, and M. Ruggieri, Rev. Mod. Phys. {\bf 86}, 509 (2014).
\bibitem{Buballa2015} M. Buballa and S. Carignano, Prog. Part. Nucl. Phys. {\bf 81}, 39 (2015).
\bibitem{Sedrakian2001} A. Sedrakian, Phys. Rev. C {\bf 63}, 025801 (2001).
\bibitem{Isayev2002} A. A. Isayev, Phys. Rev. C {\bf 65}, 031302(R) (2002).
\bibitem{Lee2018} T.-G. Lee, R. Yoshiike, and T. Tatsumi, JPS Conf. Proc. {\bf 20}, 011006 (2018).
\bibitem{Doh2006} H. Doh, M. Song, and H.-Y. Kee, Phys. Rev. Lett. {\bf 97}, 257001 (2006).
\bibitem{Zheng2015} Z. Zheng, C. Qu, X. Zou, and C. Zhang, Phys. Rev. A {\bf 91}, 063626 (2015). 
\bibitem{Zheng2016} Z. Zheng, C. Qu, X. Zou, and C. Zhang, Phys. Rev. Lett. {\bf 116}, 120403 (2016).
\bibitem{Huang2019} B. Huang, X. Yang, N. Xu, J. Zhou, and M. Gong, Phys. Rev. B {\bf 99}, 014517 (2019).
\bibitem{Kawamura2020} T. Kawamura, R. Hanai, D. Kagamihara, D. Inotani, and Y. Ohashi, Phys. Rev. A {\bf 101}, 013602 (2020).
\bibitem{Kawamura2022} T. Kawamura, R. Hanai, and Y. Ohashi, Phys. Rev. A {\bf 106}, 013311 (2022).
\bibitem{Liao2010} Y.-a. Liao, A. S. C. Rittner, T. Paprotta, W. Li, G. B. Partridge, R. G. Hulet, S. K. Baur, and E. J. Mueller, Nature (London) {\bf 467}, 567 (2010).
\bibitem{Anderson} P. W. Anderson, J. Phys. Chem. Solids. {\bf 11}, 26 (1959).
\bibitem{Clogston1} B. S. Chandrasekhar, Appl. Phys. Lett. {\bf 1}, 7 (1962).
\bibitem{Clogston2} A. M. Clogston, Phys. Rev. Lett. {\bf 9}, 266 (1962).
\bibitem{Shimahara2009} H. Shimahara, Phys. Rev. B {\bf 80}, 214512 (2009).
\bibitem{Shimahara1998} H. Shimahara, J. Phys. Soc. Jpn. {\bf 67}, 1872 (1998).
\bibitem{Ohashi2002} Y. Ohashi, J. Phys. Soc. Jpn. {\bf 71}, 2625 (2002).
\bibitem{Radzihovsky2009} L. Radzihovsky and A. Vishwanath, Phys. Rev. Lett. {\bf 103}, 010404 (2009). 
\bibitem{Radzihovsky2011} L. Radzihovsky, Phys. Rev. A {\bf 84}, 023611 (2011).
\bibitem{Yin2014} S. Yin, J.-P. Martikainen, and P. T\"{o}rm\"{a}, Phys. Rev. B {\bf 89}, 014507 (2014).
\bibitem{Jakubczyk2017} P. Jakubczyk, Phys. Rev. A. {\bf 95}, 063626 (2017).
\bibitem{Wang2018} J. Wang, Y. Che, L. Zhang, and Q. Chen, Phys. Rev. B {\bf 97}, 134513 (2018).
\bibitem{Zdybel2021} P. Zdybel, M. Homenda, A. Chlebicki, and P. Jakubczyk, Phys. Rev. A {\bf 104}, 063317 (2021).
\bibitem{Hohenberg} P. C. Hohenberg, Phys. Rev. {\bf158}, 383 (1967).
\bibitem{Mermin} N. D. Mermin and H. Wagner, Phys. Rev. Lett. {\bf 17}, 1133 (1966). 
\bibitem{note.crystal}
We note that the possibility of the first-order phase transition into a more complicated multiple-${\bm Q}$ crystalline FFLO state (which is characterized by the superfluid order parameter given in \eqref{eq.OP.FFLO}) than the FF and LO states still remains in the spatially isotropic system. In this paper, however, we only consider the revival of the second-order phase transition into the FFLO state which has been discussed in the mean-field discussions and has been precluded by considering pairing fluctuations \cite{Shimahara1998,Ohashi2002,Radzihovsky2009,Radzihovsky2011,Yin2014,Jakubczyk2017,Wang2018,Zdybel2021}. We also briefly note that, not only the second-order phase transition into the FFLO state, but also the first-order phase transition into a multiple-${\bm Q}$ crystalline FFLO state has not been observed in a spin-imbalanced Fermi atomic gas.
\bibitem{note.q1D} This instability does not contradict with the recent experiment on a quasi-one dimensional $^6$Li Fermi gas \cite{Liao2010}, where a consistent phase diagram with the existence of the FFLO state is obtained, because this system does not have continuous rotational symmetry.
\bibitem{Hidaka2015} Y. Hidaka, K. Kamikado, T. Kanazawa, and T. Noumi, Phys. Rev. D {\bf 92}, 034003 (2015).
\bibitem{Lee2015} T-G Lee, E. Nakano, Y. Tsue, T. Tatsumi, and B. Friman, Phys. Rev. D {\bf 92}, 034024 (2015).
\bibitem{Yoshiike2017} R. Yoshiike, T-G. Lee, and T. Tatsumi, Phys. Rev. D {\bf 95}, 074010 (2017). 
\bibitem{Adhikari2017} P. Adhikari, J. O. Andersen, and P. Kneschke, Phys. Rev. D {\bf 96}, 016013 (2017).
\bibitem{Pisarski2019} R. D. Pisarski, V. V. Skokov, and A. M. Tsvelik, Phys. Rev. D {\bf 99}, 074025 (2019).
\bibitem{Ferrer2020} E. J. Ferrer and V. de la Incera, Phys. Rev. D {\bf 102}, 014010 (2020).
\bibitem{Bedaque2003} P. F. Bedaque, H. Caldas, and G. Rupak, Phys. Rev. Lett. {\bf 91}, 247002 (2003).
\bibitem{Caldas2004} H. Caldas, Phys. Rev. A {\bf 69}, 063602 (2004).
\bibitem{Cohen2005} T. D. Cohen, Phys. Rev. Lett. {\bf 95}, 120403 (2005).
\bibitem{Zwierlein2006} M. W. Zwierlein, A. Schirotzek, C. H. Schunck, and W. Ketterle, Science {\bf 311}, 492 (2006).
\bibitem{Partridge2006} G. B. Partridge, W. Li, R. I. Kamar, Y. an Liao, and R. G. Hulet, Science {\bf 311}, 503 (2006).
\bibitem{Shin2006} Y. Shin, M. W. Zwierlein, C. H. Schunck, A. Schirotzek, and W. Ketterle, Phys. Rev. Lett. {\bf 97}, 030401 (2006).
\bibitem{Shin2008} Y. Shin, C. H. Schunck, A. Schirotzek, and W. Ketterle, Nature (London) {\bf 451}, 689 (2008).
\bibitem{Mitra2016} D. Mitra, P. T. Brown, P. Schau\ss{}, S. S. Kondov, and W. S. Bakr, Phys. Rev. Lett. {\bf 117}, 093601(2016).
\bibitem{Nozieres1985} P. Nozi\`{e}res and S. Schmitt-Rink, J. Low Temp. Phys. {\bf 59}, 195 (1985).
\bibitem{Tamaki2008} H. Tamaki, Y. Ohashi and K. Miyake, Phys. Rev. A {\bf 77}, 063616 (2008).
\bibitem{Chin2010} C. Chin, R. Grimm, P. Julienne, and E. Tiesinga, Rev. Mod. Phys. {\bf 82}, 1225 (2010).
\bibitem{note.trap} We briefly note that the so-called box trap has recently been realized \cite{Meyrath2005,Es2010,Gaunt2013,Mukherjee2017}, where a nearly uniform gas can be examined.
\bibitem{Meyrath2005} T. P. Meyrath, F. Schreck, J. L. Hanssen, C.-S. Chuu, and M. G. Raizen, Phys. Rev. A {\bf 71}, 041604(R) (2005).
\bibitem{Es2010} J. J. P. van Es, P. Wicke, A. H. van Amerongen, C. R\'{e}tif, S. Whitlock, and N. J. van Druten, J. Phys. B {\bf 43}, 155002 (2010).
\bibitem{Gaunt2013} A. L. Gaunt, T. F. Schmidutz, I. Gotlibovych, R. P. Smith, and Z. Hadzibabic, Phys. Rev. Lett. {\bf 110}, 200406 (2013).
\bibitem{Mukherjee2017} B. Mukherjee, Z. Yan, P. B. Patel, Z. Hadzibabic, T. Yefsah, J. Struck, and M. W. Zwierlein, Phys. Rev. Lett. {\bf 118}, 123401 (2017).
\bibitem{Dao2008} T-L. Dao, M. Ferrero, A. Georges, M. Capone, and O. Parcollet, Phys. Rev. Lett. {\bf 101}, 236405 (2008).
\bibitem{Kujawa2011} A. Kujawa-Cichy and R. Micnas, Europhys. Lett. {\bf 95}, 37 003 (2011).
\bibitem{Cichy2014} A. Cichy and R. Micnas, Ann. Phys. {\bf 347}, 207 (2014).
\bibitem{Ptok2017} A. Ptok, A. Cichy, K. Rodr\'{i}guez, and K. J. Kapcia, Phys. Rev. A {\bf 95}, 033613 (2017).
\bibitem{Wyk2018} P. van Wyk, H. Tajima, D. Inotani, A. Ohnishi, and Y. Ohashi, Phys. Rev. A {\bf 97}, 013601 (2018).
\bibitem{Ohashi2020} Y. Ohashi, H. Tajima, and P. van Wyk, Prog. Part. Nucl. Phys. {\bf 111}, 103739 (2020). 
\bibitem{Schrieffer} J. Schrieffer, {\it Theory of Superconductivity} (Addison-Wesley, New York, 1964), Chap. 7.
\bibitem{Ohashi2003}  Y. Ohashi and A. Griffin, Phys. Rev. A {\bf 67}, 063612 (2003).
\bibitem{Fukushima2007} N. Fukushima, Y. Ohashi, E. Taylor, and A. Griffin, Phys. Rev. A {\bf 75}, 033609 (2007).
\bibitem{Melo1993} C. A. R. S\'{a} de Melo, M. Randeria, and J. R. Engelbrecht, Phys. Rev. Lett. {\bf 71}, 3202 (1993).
\bibitem{Randeria} M. Randeria, in {\it Bose-Einstein Condensation}, edited by A. Griffin, D. W. Snoke, and S. Stringari (Cambridge University Press, Cambridge, UK, 1995), pp. 355-392.
\bibitem{Diener2008} R. B. Diener, R. Sensarma, and M. Randeria, Phys. Rev. A {\bf 77}, 023626 (2008).
\bibitem{He2015} L. He, H. L\"{u}, G. Cao, H. Hu, and X.-J. Liu, Phys. Rev. A {\bf 92}, 023620 (2015).
\bibitem{Bohm1949} D. Bohm, Phys. Rev. {\bf 75}, 502 (1949).
\bibitem{Ohashi1996} Y. Ohashi, and T. Momoi, J. Phys. Soc. Jpn. {\bf 65}, 3254 (1996).
\bibitem{Tajima2019} H. Tajima, T. Hatsuda, P. van Wyk, and Y. Ohashi, Sci. Rep. {\bf 9}, 18477 (2019).
\bibitem{Tempere2008} J. Tempere, S. N. Klimin, and J. T. Devreese, Phys. Rev. A {\bf 78}, 023626 (2008). 
\bibitem{Thouless1960} D. J. Thouless, Ann. Phys. {\bf 10}, 553 (1960).
\bibitem{Liu2006} X.-J. Liu and H. Hu, Europhys. Lett. {\bf 75}, 364 (2006).
\bibitem{Wardh2018} J. W\aa{}rdh, B. M. Andersen, and M. Granath, Phys. Rev. B {\bf 98}, 224501 (2018).
\bibitem{Durel2020} D. Durel and M. Urban, Phys. Rev. A {\bf 101}, 013608 (2020).
\bibitem{note2} It has been shown in the mean-field approximation that the so-called square state with the superfluid order parameter, $\Delta_i=\Delta[\cos(QR_i^x)+\cos(QR_i^y)]$, tends to be favored for a fourfold symmetric Fermi surface \cite{Shimahara2}.
\bibitem{Shimahara2} H. Shimahara, J. Phys. Soc. Jpn. {\bf 67}, 736 (1998).
\bibitem{Gubbels2013} K. Gubbels and H. Stoof, Phys. Rep. {\bf 525}, 255 (2013).
\bibitem{note.crystal2} 
If a multiple-$\bm{Q}$ crystalline FFLO state (which is not considered in the thermodynamic potential in Eq. \eqref{eq.Omega.MF}) has lower energy than the BCS state, the first-order phase transition into this crystalline FFLO state would occur. However, we do not consider this possibility in this paper.
\bibitem{Sheehy2006} D. E. Sheehy and L. Radzihovsky, Phys. Rev. Lett. {\bf 96}, 060401 (2006).
\bibitem{He2008} L. He and P. Zhuang, Phys. Rev. A {\bf 78}, 033613 (2008).
\bibitem{note.Pc1} Although we do not examine the boundary between the superfluid state and the phase separation in this paper, it is determined from the coupled superfluid number equation (\ref{eq.num}) with Eq. (\ref{eq.1st}).
\bibitem{Kashimura2012} T. Kashimura, R. Watanabe, and Y. Ohashi, Phys. Rev. A {\bf 86}, 043622 (2012).
\bibitem{Pantel2016} P-A. Pantel, D. Davesne, and M. Urban, Phys. Rev. A {\bf 90}, 053629 (2014).
\bibitem{note.U}
The value of $(k_{\rm F} a_s)^{-1}$ in the present lattice system can be estimated as follows: In the low-filling case, by expanding the atomic kinetic energy $\varepsilon_{\bm p}$ in Eq. \eqref{eq.band} around ${\bm p}=0$, the effective atomic mass $m$ can be related to the hopping parameter $t$ as $m\simeq 1/(2t)$ (when $t'=0$). For the Fermi momentum $k_{\rm F}$, we estimate it from the relation $V_{\rm FS}=4\pi k_{\rm F}^3/3$, where $V_{\rm FS}$ is given below Eq. \eqref{eq.Fermi0}. Then, for the typical parameter set $(U/(6t),n)=(0.4,0.3)$ used in this paper, $(k_{\rm F} a_s)^{-1}$ is evaluated from Eq. \eqref{eq.as} as $(k_{\rm F} a_s)^{-1} \simeq -5$, which corresponds to the weak-coupling BCS regime.
\bibitem{noteMermin} The second term in Eq. (\ref{eq.tilde.n.app}) diverges in one and two dimensions, even when one replaces ${\bm q}-{\bm Q}_\alpha^\eta$ by ${\bm q}$. This is an example of the Hohenberg-Mermin-Wagner theorem \cite{Hohenberg,Mermin}).
\bibitem{Chaikinbook} P. M. Chaikin, T. C. Lubensky, {\it Principles of Condensed Matter Physics}, Vol. 1 (Cambridge University Press, Cambridge, 1995).
\bibitem{Hornreich1975} R. Hornreich, M. Luban, and S. Shtrikman, Phys. Rev. Lett. {\bf 35}, 1678 (1975).
\bibitem{Machida1984} K. Machida and M. Fujita, Phys. Rev. B {\bf 30}, 5284 (1984).
\bibitem{Fujita1984} M. Fujita and K. Machida, J. Phys. Soc. Jpn. {\bf 53}, 4395 (1984).
\bibitem{Machida2006} K. Machida, T. Mizushima, and M. Ichioka, Phys. Rev. Lett. {\bf 97}, 120407 (2006).
\bibitem{Mizushima2007} T. Mizushima, M. Ichioka, and K. Machida, J. Phys. Soc. Jpn. {\bf 76}, 104006 (2007).
\bibitem{Mizushima2014} T. Mizushima, M. Takahashi, and K. Machida, J. Phys. Soc. Jpn. {\bf 83}, 023703 (2014) 
\bibitem{Takahashi2014} M. Takahashi, T. Mizushima, and K. Machida, Phys. Rev. B {\bf 89}, 064505 (2014).
\bibitem{Ketterle_lattice} J. K. Chin, D. E. Miller, Y. Liu, C. Stan, W. Setiawan, C. Sanner, K. Xu, and W. Ketterle, Nature {\bf 443}, 961 (2006).
\bibitem{Shimahara2021} H. Shimahara, J. Phys. Soc. Jpn. {\bf 90}, 044706 (2021).
\bibitem{Matsuda2007} Y. Matsuda and H. Shimahara, J. Phys. Soc. Jpn. {\bf 76}, 051005 (2007).
\bibitem{Son2006} D. T. Son and M. A. Stephanov, Phys. Rev. A {\bf 74}, 013614 (2006).
\bibitem{Bulgac2008} A. Bulgac and M. M. Forbes, Phys. Rev. Lett. {\bf 101}, 215301(2008).
\bibitem{Tuzemen2020}  B. T\"{u}zemen, P. Kukli\'{n}ski, P. Magierski, and G. Wlaz\l{}owski, Acta Phys. Pol. B {\bf 51}, 595 (2020).
\end{thebibliography}
\end{document}